\documentclass[twocolumn]{aastex631}

\newcommand{\degree}{$^\circ$ }
\newcommand{\degW}{$^\circ$W }
\newcommand{\degN}{$^\circ$N }
\newcommand{\degS}{$^\circ$S }
\newcommand{\arcsecs}{\arcsec \ }
\newcommand{\um}{\textmu m }
\newcommand{\microns}{\textmu m}
\usepackage{chemformula}
\usepackage{chemfig}
\usepackage{float}
\usepackage{amssymb}

\received{March 6, 2025}
\revised{May 11, 2025}
\accepted{May 28, 2025}
\submitjournal{PSJ}

\graphicspath{{.}{figures/}}
\begin{document}

\title{Spectroscopic Mapping of Callisto with HST/STIS and Implications for its Surface Composition}

\correspondingauthor{M. Ryleigh Davis}
\email{rdavis@caltech.edu}

\author[0000-0002-7451-4704]{M. Ryleigh Davis}
\affiliation{Division of Geological and Planetary Sciences, California Institute of Technology, Pasadena, CA 91125, USA}

\author[0000-0002-0767-8901]{Samantha K. Trumbo}
\affiliation{Department of Astronomy \& Astrophysics, University of California San Diego, La Jolla, CA 92093, USA}

\author[0000-0002-8255-0545]{Michael E. Brown}
\affiliation{Division of Geological and Planetary Sciences, California Institute of Technology, Pasadena, CA 91125, USA}

\author[0000-0003-4778-6170]{Mathew Belyakov}
\affiliation{Division of Geological and Planetary Sciences, California Institute of Technology, Pasadena, CA 91125, USA}

\begin{abstract}
We present global, spatially resolved ultraviolet-visible spectra of Callisto obtained with HST/STIS and explore possible compositions of Callisto’s surface material. We map the strength of a widespread downturn toward the near-UV and the NIR spectral slope from 700 to 1000 nm, which varies from slightly blue (reflectance decreasing from 700 to 1000 nm) to red (reflectance increasing) across Callisto's surface. Globally, bright water-ice-rich regions tend to have neutral or blue NIR slopes and a shallower near-UV downturn, while darker material is associated with red NIR slopes and stronger near-UV absorption. Broad absorptions near 820 and 930 nm are spatially correlated with the Asgard and Valhalla impact basins and may be associated with iron-bearing silicates. An absorption edge near 275 nm maps primarily to Callisto's trailing hemisphere, and a 320 nm absorption most prevalent within and surrounding Asgard and Valhalla may be related to organics. We report two new absorption features near 230 and 450 nm which might be attributed to irradiated \ch{NaCl}. We find little evidence for sulfur-bearing species at UV-visible wavelengths and suggest that a 280 nm band seen only in leading/trailing hemisphere ratio spectra and previously attributed to \ch{SO2} is better explained as a consequence of dividing the unrelated 320 nm leading hemisphere band by the trailing-hemisphere 275 nm absorption edge. Spatial variations in spectral features suggest that Callisto's dark material composition varies regionally, reflecting a mix of endogenic and exogenic sources and radiolytic alteration.
\end{abstract}.

\keywords{Galilean satellites(627) --- Callisto(2279) --- Surface composition(2115) --- Natural satellite surfaces (2208) --- Ultraviolet spectroscopy(2284)
}

\section{Introduction} \label{sec:intro}

Callisto, the outermost of Jupiter's Galilean moons, has an ancient ($\sim$4-4.5 Ga) surface dominated by impact craters and no evidence for any past endogenic processing, volcanism, tectonics, or other surface-interior interactions which are widespread on Io, Europa, and Ganymede \citep{greeley2000_GalileoViewsGeologya, moore2004callisto, bottke2024bombardment}. Callisto has the lowest albedo and the highest proportion of dark material among the four Galilean moons \citep{moore2004callisto}. Its surface is also the most degraded, and sublimation-driven landform modification and mass-wasting effects result in a well-segregated surface. Bright high-albedo regions consist of relatively pure frost emplaced on topographic highs, formed when exposed \ch{H2O} ice sublimates under daytime heating ($\sim$165 K) and re-deposits on crater rims and peaks \citep{moore1999_MassMovementLandform, howard2008_SublimationdrivenErosionCallisto}. A smooth blanket of fine-grained dark (non-icy) material fills topographic lows like crater floors and inter-crater depressions \citep{moore2004callisto}. This dark material may be largely composed of lag deposits left behind after the \ch{H2O} (and possibly \ch{CO2}) ice sublimates out of endogenic ice/rock mixtures which were excavated from the subsurface by large impactors \citep[e.g.][]{moore1999_MassMovementLandform, moore2004callisto, howard2008_SublimationdrivenErosionCallisto, white2016modeling}, or it may be dominated by dust delivered from the irregular satellites via Poynting–Robertson drag \citep{bottke2013black, chen2024_IrregDust}. Most likely, Callisto's dark material represents a mix of endogenic lag deposits and delivered dust, but their relative contributions remain uncertain. In addition to this dark material, Callisto's surface is composed primarily of water ice, small amounts of CO$_2$, and possibly sulfur-bearing species \citep{noll1997_DetectionSO2Callisto, moore2004callisto, hibbitts2002co2, cartwright2020_EvidenceSulfurbearingSpecies, cartwright2024_RevealingCallistoCarbonrich}.

Despite nearly 50 years of spectroscopic study, the specific mineralogical composition of Callisto's dark material remains largely unknown. Several lines of evidence, including absorption features detected with Galileo/NIMS (0.7-5.2 \microns, R$\sim$40-200), ground-based observations, and recent JWST/NIRSpec (G395H; 2.85-5.35 \microns, R$\sim$2700) data have been interpreted as arising from iron- and Mg-OH bearing silicates, carbon bearing species, and organic molecules, perhaps suggesting a composition similar to primitive asteroids and carbonaceous chondrites \citep{calvin1991_ModelingReflectanceSpectrum, calvin1993spectral, mccord1998_NonwatericeConstituentsSurfacea, mccord1997_OrganicsOtherMolecules, cartwright2024_RevealingCallistoCarbonrich}. However, the identities of many non-water ice absorptions between 1 and 5 \um remain uncertain, and the literature is often contradictory, with various types of material suggested to explain each feature. Some authors suggest that several of these NIR bands are consistent with carbonates and C-H and C-N stretch features from organics \citep{mccord1998_NonwatericeConstituentsSurfacea, johnson2004_MagnetosphericPlasmadrivenEvolution, cartwright2024_RevealingCallistoCarbonrich}, while others claim they are more consistent with sulfur-bearing species \citep{mccord1997_OrganicsOtherMolecules, cartwright2020_EvidenceSulfurbearingSpecies}, and still others suggest that at least some of these features may be associated with ammonia or NH\textsubscript{4}-bearing minerals \citep{calvin1993spectral} or Na-bearing species like Na-carbonates \citep{johnson2004radiationchapter}, Na-oxides (e.g. \ch{Na2C2O4}), or Na-sulfates (e.g. \ch{Na2SO4}) \citep{cartwright2020_EvidenceSulfurbearingSpecies, cartwright2024_RevealingCallistoCarbonrich}, all of which have different implications for Callisto's formation and history. 

Existing observations of Callisto at UV and visible wavelengths have led to a number of valuable insights, despite the somewhat sparse availability of data. Callisto displays a significant hemispheric albedo dichotomy, where its trailing hemisphere (270\degW) is brighter than its leading hemisphere (90\degW), which is opposite of the pattern seen on Europa and Ganymede \citep{nelson1987_SpectralGeometricAlbedos, hendrix2005icy}. This dichotomy may result directly from the preferential delivery of dark material from Jupiter's prograde irregular satellites to the leading hemisphere \citep{bottke2013black, chen2024_IrregDust}, or from weathering processes thought to alter Callisto's surface including micrometeoroid gardening and radiolytic alteration from magnetospheric charged particle bombardment \citep{buratti1991ganymede, johnson2004radiationchapter, bottke2013black}. A strong opposition effect on the leading hemisphere, where it becomes nearly equally as bright as the trailing side when viewed at 0\degree solar phase angle, may arise from smaller grain sizes resulting from the higher total bombardment of irregular satellite dust onto the leading hemisphere \citep{buratti1991ganymede, calvin1993spectral}. 

A UV absorption feature near 280 nm, seen in the spectral ratio of Callisto's leading to trailing hemisphere, has been attributed to \ch{SO2} possibly related to neutral sulfur from Io delivered to the the leading/sub-Jovian hemisphere or due to enhancement of endogenic or delivered sulfur by micrometeoroid bombardment \citep{lane1997so2, noll1997_DetectionSO2Callisto}. The presence of \ch{SO2} on Callisto was initially supported by near-IR absorptions at 3.88 and 4.02 \um seen in some Galileo/NIMS observations and suggested to arise from an S-H stretch \citep{mccord1997_OrganicsOtherMolecules} and \ch{SO2} \citep{hibbitts2000_DistributionsCO2SO2}, respectively. However, more recent analyses have questioned the identification of all three of these absorptions \citep{johnson2004_MagnetosphericPlasmadrivenEvolution, hendrix2008_CallistoNewInsights, cartwright2020_EvidenceSulfurbearingSpecies}. Whether or not \ch{SO2} or endogenically sourced sulfur-bearing species are present on Callisto remains an open question, although some amount of Iogenic sulfur is likely delivered by the Jovian magnetosphere \citep{cooper2001energetic, liuzzo2019energetic}. 

Generally, Callisto's UV slope tends to redden at lower latitudes, which \citet{hendrix1999_HYDROGENPEROXIDEICY} suggest may be due to the presence of \ch{H2O2}. However, Callisto lacks the 3.5-\um \ch{H2O2} absorption seen on Europa and Ganymede \citep{carlson1999_HydrogenPeroxideSurface, hand2013peroxide, trumbo2019_H2O2ChaosTerrain, trumbo2023_HydrogenPeroxideGanymede, wu2024europa}, calling this identification into question. Analysis of Galileo/UVS data ($\sim$160-320 nm, R$\sim$150-230) revealed the presence of a possible absorption feature at high southern latitudes with a band center somewhere long-wards of 320 nm (where the UVS data ended) which has been attributed to organic species \citep{hendrix2008_CallistoNewInsights}. \citet{hendrix2008_CallistoNewInsights} hypothesize that this species is progressively weathered at lower latitudes by charged particle or UV irradiation, producing carbonized organics which absorb strongly in the far-UV and potentially explaining both the observed variation in the UV spectral slope and the apparent 280 nm absorption seen in leading/trailing ratio spectra. In this case, much of Callisto's UV spectral behavior is explained by organic species which have undergone varying amounts of radiolytic alteration rather than the previously suggested \ch{H2O2} and \ch{SO2}, perhaps suggesting a largely carbon dominated surface chemistry.

While existing data sets remain consistent with the presence of hydrated silicate minerals, carbon-bearing species, CH- or CN- bearing organics, and S-bearing species within Callisto's dark material, they have proven insufficient for definitive identification and evidence for S-bearing species has become increasingly sparse. Previous analyses at UV and visible wavelengths have relied on full-disk UV-visible spectroscopy, which lack spatial detail \citep[e.g.][]{mcfadden1980_VisibleSpectralReflectance, nelson1987_SpectralGeometricAlbedos, hendrix2005icy}; spatially resolved UV spectra with limited spatial and wavelength coverage ($\sim$160-320 nm) 
\citep{hendrix2008_CallistoNewInsights}; or high-spatial-resolution imagery, which lacks spectral detail \citep{denk1999galileo, greeley2000_GalileoViewsGeologya}. To help expand our understanding of the spectral properties of Callisto's surface, we used the Space Telescope Imaging Spectrograph aboard the Hubble Space Telescope (HST/STIS; $\sim$200-1000 nm, R$\sim$600-1000) to obtain the first near-global, spatially resolved UV-visible wavelength reflectance spectra of Callisto. We use this new data set to identify and map absorption features across Callisto's surface and discuss compositional implications in the context of potential endogenic materials, delivery of exogenic species, and alteration via charged particle bombardment.

\section{Observations and Data Reduction} \label{sec:data}
\begin{table*}
\caption{Table of HST/STIS Observation Details}\label{tab:obs}
\begin{tabular}{cccccccc}
    \hline\\[-4mm] \hline
    Date&Time&Central&Central&Angular&Gratings&Slit&Exposure\\
    (UT)&(Start-End)&Lon.&Lat.&Diameter&Used&Width&Time (s)\\ 
    \hline
    2020-08-22 & 04:47-05:20 & 47\degree W & -1.08\degree & 1.53\arcsec & G230L & 0.2\arcsec & 155.0\\
    2020-08-26 & 05:43-06:16 & 135\degree W & -1.09\degree & 1.52\arcsec & G230L & 0.2\arcsec & 155.0\\
    2020-10-19 & 17:17-17:51 & 229\degree W & -1.01\degree & 1.29\arcsec & G230L & 0.2\arcsec & 155.0\\
    2020-09-20 & 09:31-10:04 & 318\degree W & -1.07\degree & 1.41\arcsec & G230L & 0.2\arcsec & 155.0\\
    2020-09-07 & 13:20-13:53 & 41\degree W & -1.08\degree & 1.46\arcsec & G430L & 0.1\arcsec & 42.0\\
    2020-10-10 & 00:00-23:59 & 41\degree W & -1.03\degree & 1.32\arcsec & G750L & 0.1\arcsec & 75.0\\
    2020-10-15 & 08:26-08:55 & 135\degree W & -1.02\degree & 1.31\arcsec & G430L & 0.1\arcsec & 42.0\\
    2020-10-15 & 09:52-10:30 & 136\degree W & -1.02\degree & 1.31\arcsec & G750L & 0.1\arcsec & 70.0\\
    2021-05-26 & 05:42-06:12 & 221\degree W & 0.65\degree & 1.37\arcsec & G430L & 0.1\arcsec & 42.0\\
    2020-10-19 & 09:13-09:51 & 221\degree W & -1.01\degree & 1.29\arcsec & G750L & 0.1\arcsec & 70.0\\
    2020-05-10 & 03:15-03:45 & 314\degree W & -1.1\degree & 1.41\arcsec & G430L & 0.1\arcsec & 42.0\\
    2020-05-10 & 04:39-05:18 & 315\degree W & -1.1\degree & 1.41\arcsec & G750L & 0.1\arcsec & 75.0\\
    \hline
\end{tabular}
\end{table*}

We present global, spatially resolved ultraviolet and visible wavelength spectroscopy of Callisto's surface taken with HST/STIS. The data were acquired in the first-order spectroscopy mode with the G230L (157 - 318 nm), G430L (290 - 570 nm) and G750L (524 - 1027 nm) gratings across 10 visits between May 2020 and May 2021. UV observations with the G230L grating were taken with a slit width of 0.2\arcsecs and used the STIS MAMA detector, while the two visible gratings (G430L and G750L) used a 0.1\arcsecs slit and the STIS CCD. 

The observations are summarized in Table \ref{tab:obs}. We downloaded flux- and wavelength-calibrated x2d files from the MAST Archive produced by the standard HST STIS calibration pipeline (calstis). Data taken with the G750L grating showed substantial fringes at the longest wavelengths, so we reprocessed the data using the calstis pipeline with the additional defringing routines turned on. We then extracted individual spectra from the x2d images, with a pixel scale of 0.025\arcsecs and 0.05\arcsecs for the UV and visible gratings, respectively. This corresponds to a spatial resolution of $\sim$80 and $\sim$160 km near the center of the disk on Callisto, which is comparable to the diffraction-limited spatial resolution of HST near 250 nm (80 km) and 600 nm (192 km), respectively.  To convert each spectrum to units of reflectance, we divided by a solar spectrum which was smoothed and binned to match the spectral resolution and sampling of the observations. For the G230L grating, we used solar spectra from the Total and Spectral Solar Irradiance Sensor (TSIS-1) Spectral Irradiance Monitor (SIM) on the International Space Station acquired on the same date as each observation and downloaded via the LASP Interactive Solar Irradiance Datacenter (LISIRD)\footnote{https://lasp.colorado.edu/lisird/}. For the two visible wavelength gratings, we used the ASTM E490 zero-airmass solar reference spectrum\footnote{https://www.nrel.gov/grid/solar-resource/spectra-astm-e490.html}.

The reprocessed G750L spectra exhibited multiple artifacts as a result of the defringing process as well as an artificial dropping of the reflectance towards longer wavelengths as a result of slit losses due to the broadening of the point-spread function with increasing wavelength, as was also noted in STIS observations of Mars, Europa and Io \citep{bell2007high, trumbo2020endogenic, trumbo2022_SpectroscopicMappingIo}. We simultaneously correct for these wavelength-dependent slit losses and defringing artifacts following the method described in \citet{trumbo2022_SpectroscopicMappingIo}. We fit a spline curve to the ground-based leading-hemisphere spectrum of Callisto from \citet{spencer1995charge}, combined with the lower spectral resolution data from \citet{clark1980_GalileanSatellitesNew} to fill in missing wavelengths above $\sim$780 nm. We then construct a disk-integrated spectrum of Callisto's leading hemisphere, centered at 70\degW to match the viewing geometry of \citet{spencer1995charge}. We multiply each individual HST spectrum by the ratio of the spline fit to the ground based data and the constructed full-disk HST spectrum. This allows for the correction of the slit losses and defringing artifacts while preserving pixel-to-pixel variations and the general spectral shape within the data.

\begin{figure*}[ht!]
\includegraphics[width=\textwidth]{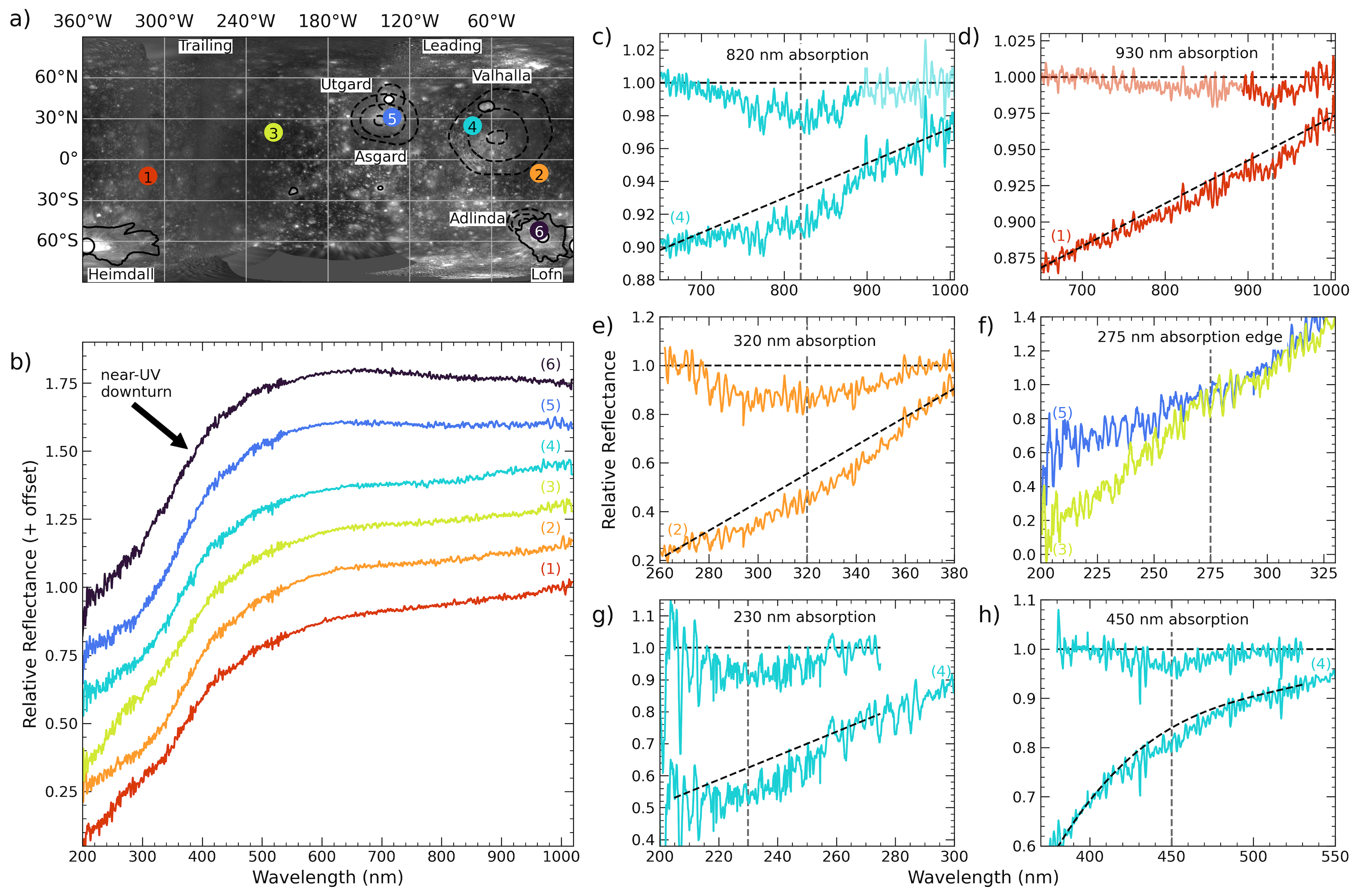}
\caption{Example HST spectra from 200 to 1000 nm of a few representative regions on Callisto. The location of each spectrum is indicated by a numbered circle on the map (a) with a corresponding color to the plotted spectrum (b). Each spectrum is calculated by averaging the UV and visible pixels which intersect a circle with a radius of 2\degree about the point of interest, and combining the resulting UV and visible spectra. Spectra are normalized to unity at 1000 nm and offset by increments of 0.15 for clarity. These spectra are generally smoothly varying, and uncertainties can be estimated from the scatter in the data. (c-h) show many of the individual spectral features discussed in this paper, with band centers indicated by the vertical dashed gray lines. Colors indicate the corresponding spectrum from panel b used to highlight each individual spectral feature. Polynomial continuum fits are shown with black dashed lines and continuum divided features are included above each spectrum. In panel f, the Asgard spectrum (5, blue) is included to highlight the difference between a typical Callisto spectrum that does not have a 275 nm absorption edge and one that does (3, yellow). The large ringed basins Valhalla, Asgard, Utgard, and Adlinda are outlined with black dashed lines and labeled on the map in panel (a) along with the bright young impact craters Heimdall and Lofn and their ejecta. The basemap is the high-resolution USGS global mosaic created from Voyager and Galileo images \citep{usgs2013callisto}.}
\label{fig:roi}
\end{figure*}

We calculated the geographic coordinates of each extracted pixel on Callisto using the phase, angular size, and pole orientation at the time of the observation obtained from JPL Horizons. We found that the telemetry information contained in the FITS headers was not sufficiently accurate to determine the geographic locations of each pixel, particularly for the observation taken on 2021-05-26, where the HST target acquisition appears to have partially failed. We therefore reconstructed a 3d image from each of the individual slit observations for a given grating and viewing geometry, summed across the wavelength axis to create a 2d average image, and fit a 2d Gaussian to the reconstructed image to identify the center of the disk. We used this disk center and the slit-offset information contained in each fits header to calculate the corresponding latitude and longitude of each spatial pixel. We then combined the two visible gratings (G430L and G750L) to create a full visible wavelength spectrum (300 - 1000 nm) for each spatial pixel.  For the 45\degW, 135\degW, and 315\degW  viewing angles, the two visible gratings were taken close enough in time and with the same slit orientation and step pattern that the resulting pixel locations on Callisto were very similar and we simply matched up pixels which overlapped in area by more than $\sim$90\%. For the two gratings at $\sim$225\degW viewing geometry, however, the observations were taken nearly 7 months apart and combining the full visible wavelength region required more care. We interpolated the spatial pixels in the G750L grating to match those of the G430L grating by computing the average of the spectra of G750L pixels which overlap a given G430L pixel, weighted by the area of the overlapping region. For all viewing angles, we computed an additional offset, which was applied to the G750L spectra so that the median values of the spectra at the overlapping wavelengths from the two gratings are equal. This ensured that there were no jumps or discontinuous regions in the combined visible wavelength reflectance spectra. We generally consider the UV and visible spectra separately, at their native resolutions, except when otherwise noted. In this case, we degrade the spatial resolution of the visible data to match that of the UV observations by resampling the visible data at the locations of the UV spatial pixels. We do not apply a specific photometric correction because we analyze normalized spectra. However, pixels whose center coordinates were farther than 60\degree  from the disk center were removed in order to minimize photometric effects associated with high incidence angles where deviations from ideal photometric behavior may become significant, except for the case where there were no other data associated with that spatial region.

\section{Definition of Spectral Features}

Figure \ref{fig:roi} shows a selection of representative spectra on Callisto's surface. Immediately obvious in all of the spectra is a sharp decrease in reflectance from the visible to near-UV wavelengths, consistent with previous ground-based disk-averaged observations \citep[e.g.][]{johnson1970_GalileanSatellitesSpectral, clark1980_GalileanSatellitesNew}. Indeed, all of the Galilean satellites have a spectrum with a band edge near 500 nm, interpreted as the edge of a broad and strong UV absorption \citep[e.g.][]{mcfadden1980_VisibleSpectralReflectance, calvin1995_SpectraIcyGalileanc}. This near-UV downturn appears to be present globally on Callisto, although its strength varies across the surface, and the wavelength at which the downturn starts shifts from $\sim$ 500 to 650 nm. All of the Callisto spectra have red slopes in the near-UV (200 - 350 nm), consistent with \citet{hendrix2008_CallistoNewInsights}, and a steep red slope from $\sim$350 - 500 nm with an apparent slope break near 415 nm. Most of the spectra have a neutral or moderately red NIR slope from 700 to 1000 nm, although some pixels have a slightly blue slope beyond 700 nm.

As is shown in Figure \ref{fig:roi} (c-h), there are a number of discrete absorption features seen in the Callisto spectra. There are two broad and shallow absorptions centered at 820 and 930 nm which combine to form a broader band from  $\sim$ 700 to 990 nm in many pixels. This combined band is consistent with the weak $\sim$ 750 to 950 nm absorption reported in the ground-based spectrum of \citet{clark1980_GanymedeEuropaCallisto}. Some of the Callisto spectra also display an absorption from 275 - 360 nm, with a band center near 320 nm. Figure \ref{fig:roi}e shows an example spectrum with a clear 320 nm absorption. It is possible that this feature is the same as the $\gtrapprox$320 nm absorption inferred by \citet{hendrix2008_CallistoNewInsights}, see Section \ref{sec:320nm} for more detailed discussion. Some regions on Callisto's trailing hemisphere show a steep absorption edge starting at $\sim$275 nm and quickly dropping towards shorter wavelengths (see spectra 1 and 3 in Figure \ref{fig:roi}b and the zoom-in in panel f). Additionally, many of the Callisto spectra also show two previously unreported absorption features centered near 230 nm and 450 nm, as seen in Figure \ref{fig:roi} g and h. These features are superimposed on the much stronger 275 nm absorption edge and near-UV downturn, respectively.

\citet{noll1997_DetectionSO2Callisto} noted that the spectral ratio of Callisto's leading to trailing hemisphere produces a broad absorption feature centered near 280 nm, which we are able to reproduce with our data set. However, the width of the ratio feature is wider than that assumed by \citet{noll1997_DetectionSO2Callisto}, whose data only extended to 330 nm (see Section \ref{sec:libcomp_280nm} for a more detailed discussion of the spectra), and we see no unambiguous evidence of a UV absorption feature centered at 280 nm in the spectra of any of the individual spatial pixels without performing such a ratio.

In the following sections, we explore the spatial distributions of the near-UV downturn, NIR (700 - 1000 nm) slope, 930 nm, 820 nm, 450 nm, 320 nm and 230 nm absorption features, 280 nm ratio feature, and the 275 nm absorption edge and compare the spectra with existing libraries of laboratory spectra to explore possible compositions of Callisto's non-ice material. We then consider the spatial distributions and plausible identifications of the various bands and propose a consistent conceptual model for the composition of Callisto's dark material with contributions of endogenic, exogenic, and radiolytically processed species. Finally, we compare the average leading and trailing hemisphere spectra of Callisto with equivalent HST/STIS spectra of the other Galilean satellites - Io, Europa and Ganymede and discuss implications for the Galilean system.

\begin{figure*}[hp!]
\centering
\includegraphics[width=0.9\textwidth]{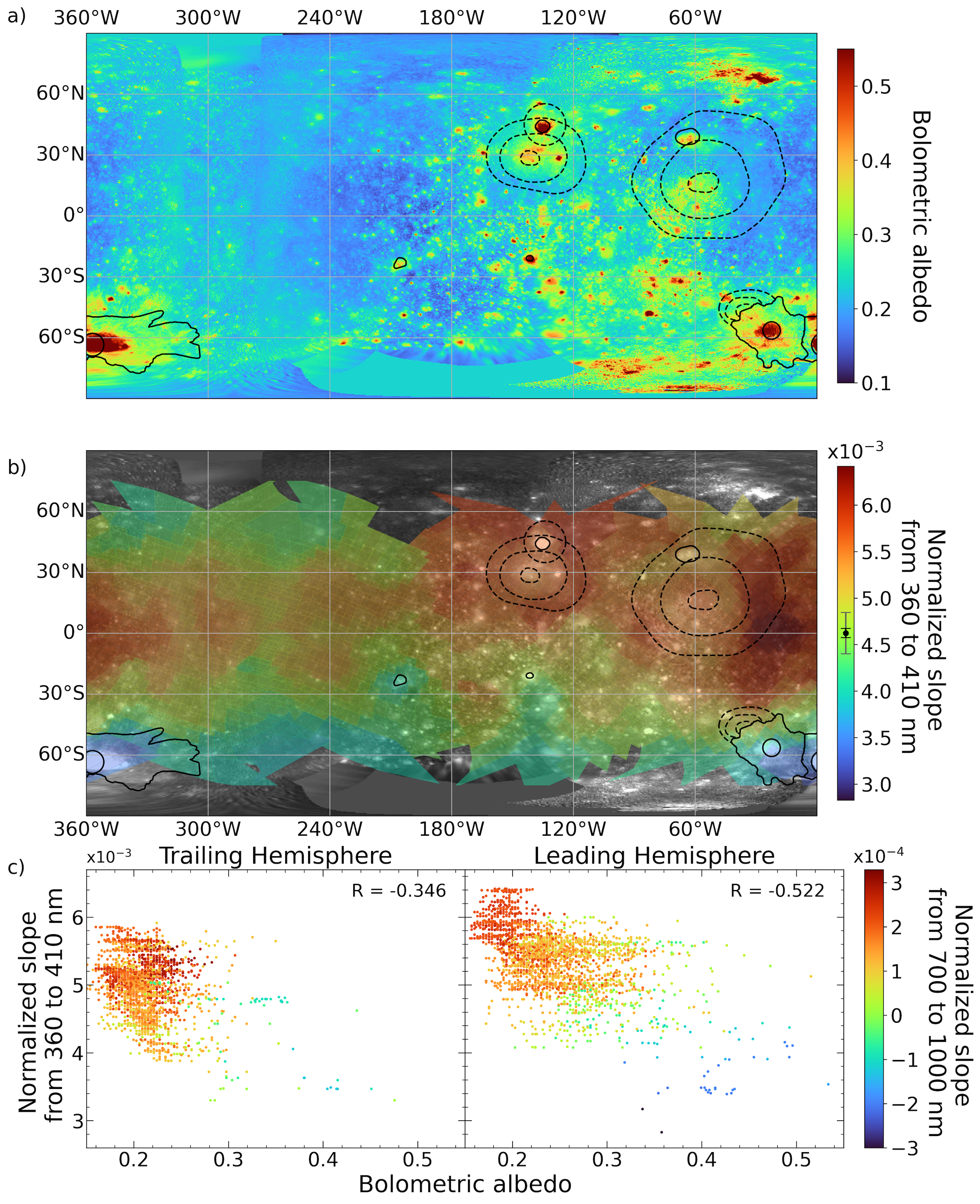}
\caption{Maps showing Callisto's bolometric albedo \citep[a; from][]{camarca2023_ThermalPropertiesLeading} compared with the spectral slope from 360 to 410 nm (b), which measures the strength of the near-UV downturn (redder slope = stronger downturn). The per-pixel estimated 1$\sigma$ error on the 360 to 410 nm slope ranged from $\pm$4.9$\times$10$^{-5}$ to $\pm$2.2$\times$10$^{-4}$, and the minimum (black) and maximum (grey) estimated error is shown on the colorbar. On both maps, the large ringed basins Valhalla (56\degW, 16\degN), Asgard (140\degW, 32\degN), Utgard (134\degW, 45\degN), and Adlinda (36\degW, 48\degS) are outlined with black dashed lines. For Valhalla, Asgard, and Adlinda, the central, inner, and outer rings are shown, based on the geologic map from \citet{greeley2000_GalileoViewsGeologya}. A few young, bright impact features are also indicated in solid black, along with Heimdall (357\degW, 65.5\degS) and Lofn (22\degW, 56.5\degS) craters and their relatively extensive ejecta. The scatter plots (c), produced by degrading the spatial resolution of the albedo map to match the HST/STIS observations, reveal the moderate correlation between the bolometric albedo and near-UV downturn on Callisto's leading and trailing hemispheres and the Pearson correlation coefficient (R) is shown on each plot. The colorbar indicates the NIR slope (700-1000 nm) mapped in Figure \ref{fig:map_700_1000nm}. }
\label{fig:uvdropoff}
\end{figure*}

\begin{figure*}[ph!]
\includegraphics[width=\textwidth]{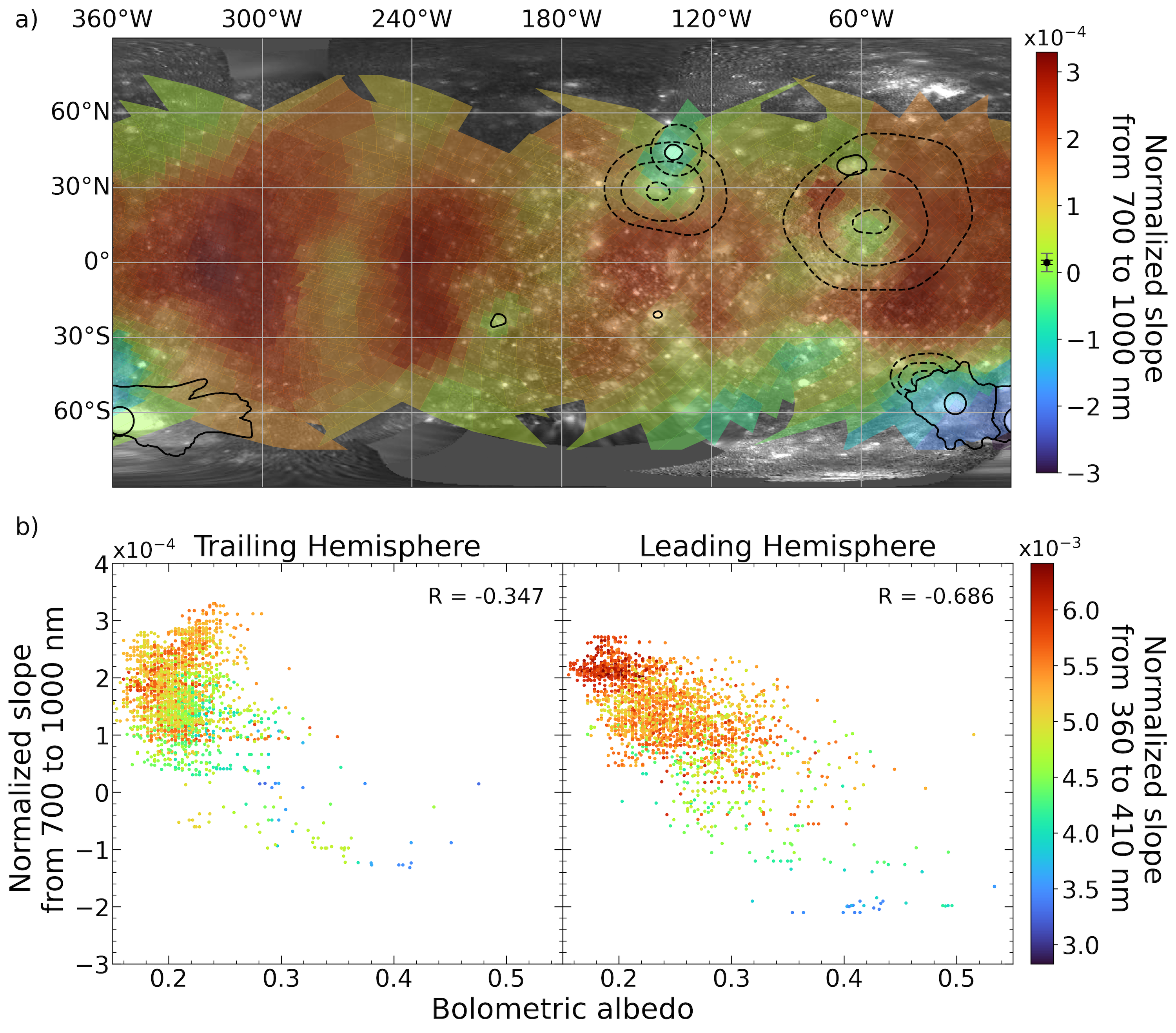}
\caption{(a) Map showing the NIR spectral slope from 700 to 1000 nm. The per-pixel estimated 1$\sigma$ error on the slope ranged from $\pm$3.3$\times$10$^{-6}$ to $\pm$1.5$\times$10$^{-5}$, and the minimum (black) and maximum (grey) estimated error is shown on the colorbar. Most of the surface has a neutral or red NIR slope except for the regions at southern latitudes and near Heimdall and Lofn that show an abundance of icy material and the central zones of the Valhalla, Asgard, and Utgard ringed basins which have slightly blue slopes. (b) Scatter plots demonstrating the moderate correlation between the albedo and NIR slope, especially on the leading hemisphere. Pixels with the reddest NIR slopes tend to be dark, but not all dark material is particularly red. A moderate correlation between NIR red and UV red material is seen in the dark material on the leading hemisphere, but such a relationship is not as apparent on the trailing hemisphere.}
\label{fig:map_700_1000nm}
\end{figure*}

\section{Spectral Maps}\label{sec:maps}

We map the strength of Callisto's absorption features with one of two methods. For the near-UV downturn, where the full width of the absorption extends beyond the observed wavelength range, as well as the NIR (700 - 1000 nm) slope, we measure the strength of the feature by fitting a line to the data after normalizing at the long wavelength edge, excluding wavelengths that may be strongly affected by other overlapping features. We estimate 1$\sigma$ uncertainties on the measured slope values for each spatial pixel using the residual variance of the linear polynomial fit and the scatter in the spectral values, using standard linear regression error propagation \citep{bevington2003}. For the near-UV downturn, we measure the slope between 360 to 410 nm in order to avoid interference from the 320 nm and 450 nm features and normalize at 410 nm. We normalize at 1000 nm and measure the slope between 680 and 1000 nm (excluding 700 - 930 nm where the 820 and 930 nm bands are) for the NIR slope, although the effect of the weak 820 and 930 nm bands is minimal and slopes computed across the entire 700 to 1000 nm range produce qualitatively similar maps. The results are mapped in Figure \ref{fig:uvdropoff} and Figure \ref{fig:map_700_1000nm}, respectively. We find that small changes to the wavelength bounds used in the linear fits produce consistent results for these mapped slopes.

For the 280 nm ratio, 320 nm, and combined 820 and 930 nm features, we map the band strengths using the integrated band area. We fit a linear continuum to the regions surrounding the absorption and produce a continuum divided spectrum. We then integrate the residual absorption across the full feature to calculate a band area or equivalent width and estimate the 1$\sigma$ uncertainty for each spatial pixel by taking the standard deviation of the normalized spectrum within the feature as a constant 1$\sigma$ uncertainty and computing the trapezoidal error propagation for the band area integration as described in \citet{bevington2003}. For the overlapping 820 and 930 nm bands, we calculate a single combined band strength by fitting a linear continuum from 660 to 995 nm (excluding 695 - 975 nm) and integrating across the combined band width from 710 to 980 nm. We manually assess each fit and make small changes to these bounds when necessary, which is especially important for pixels which show an apparent downturn near 1000 nm possibly associated with the 1.04 \um absorption feature noted in some ground-based observations \citep[e.g.][]{clark1980_GalileanSatellitesNew}. For the 320 nm feature, we fit the continuum from 262 to 370 nm (excluding 275 - 360 nm) and integrate across the band from 275 to 360 nm. For the 280 nm ratio feature, we first construct an average spectrum within 5\degree of the trailing hemisphere apex (270\degW, 0\degN) and divide each Callisto spectrum by this trailing hemisphere average before fitting a linear continuum from 230 to 405 nm (excluding 235 - 400 nm) and integrating the residual feature from 235 to 400 nm. These calculated band areas are mapped in Figures \ref{fig:820nm_930nm}, \ref{fig:320nm}, and  \ref{fig:280nm}.

Finally, for the 275 nm absorption edge, it is difficult to determine a consistent UV spectral continuum for each pixel because the feature extends beyond the observed wavelength range, there is increased noise at the short-wavelength end of the spectra, and additional overlapping absorption features apparently modify Callisto's UV spectrum (e.g. the 230 nm feature). We therefore apply both approaches for measuring the band strength and compare the resulting maps. We measure the slope between 230 and 270 nm after normalizing at 270 nm, with the results shown in Figure \ref{fig:275nm_edge}a. To measure the band area, we first construct a base spectrum from Callisto spectra interpreted to contain little or no sign of the 275 nm absorption edge by averaging five pixels with the smallest (nearly neutral) 230 to 270 nm slopes. The five pixels used in this mean are nearby one another and centered at (141\degree W, 42\degree N), (154\degree W, 39\degree N), (150\degree W, 45\degree N), (145\degree W, 51\degree N), and (144\degree W, 49\degree N). We then divide each Callisto spectrum, normalized to unity at 280 nm, by this base spectrum and integrate the residual feature from 220 to 275 nm. We exclude wavelengths shorter than 220 nm because of to the significantly increased noise at the short-wavelength end of the spectra, as well as the presence of a possible absorption shoulder short-wards of $\sim$220 nm in some of the pixels. The results are mapped in Figure \ref{fig:275nm_edge}b.

In the following subsections, we map the strength of each spectral feature and discuss their spatial distributions. We report the minimum and maximum estimated uncertainties across all of the spatial pixels as error bars on each figure's corresponding colorbar and caption.

\subsection{Near-UV Downturn}\label{sec:downturn}

Figure \ref{fig:uvdropoff}a shows the bolometric albedo from \citet{camarca2023_ThermalPropertiesLeading} for comparison with the strength of the near-UV downturn mapped in Figure \ref{fig:uvdropoff}b. Generally, the leading hemisphere has a stronger near-UV absorption than the trailing hemisphere, especially at northern latitudes. The leading hemisphere southern low latitudes, where there is an abundance of bright material, show a relatively weak near-UV downturn. The southern low latitudes on the leading/anti-Jovian hemisphere is also where \citet{cartwright2024_RevealingCallistoCarbonrich} find the strongest 4.57 \um band strengths, which is tentatively attributed to CN-bearing organics.  The trailing hemisphere mid-latitudes show a stronger near-UV downturn than the high-latitudes, which have a similar strength as the leading hemisphere low southern latitudes. However, as seen in Figure \ref{fig:uvdropoff}a, the trailing hemisphere high-latitudes do not have an abundance of brighter material like the leading hemisphere low southern latitudes. The bright young impact features Heimdall (357\degW, 65.5\degS) and Lofn (22\degW, 56.5\degS) and their ejecta show the shallowest measured downturn, while the low albedo region to the east of Valhalla have the strongest measured near-UV downturn.

Figure \ref{fig:uvdropoff}c shows a scatter plot of the bolometric albedo versus our proxy for the strength of the near-UV downturn for the leading and trailing hemispheres with the colorbar indicating the NIR (700 to 1000 nm) slope mapped in Figure \ref{fig:map_700_1000nm}. The bolometric albedo of each HST spatial pixel is determined by calculating the average albedo of overlapping pixels from the higher spatial resolution map of \citet{camarca2023_ThermalPropertiesLeading}, thereby degrading the resolution to match that of the HST observations. Generally, pixels with the shallowest near-UV downturns tend to have higher albedos and regions with the strongest downturns have low albedos, explaining the moderate negative correlation. However, relative differences in the albedo of the dark material do not correlate strongly with the strength of the near-UV downturn, aside from having a generally steeper slope than high albedo pixels. Additionally, while the bright young impact features Heimdall and Lofn tend to have a much shallower near-UV downturn compared with the Valhalla (56\degW, 16\degN) and Asgard (140\degW, 32\degN) impact basins, the strength of the near-UV absorption within these geologic features is not noticeably different from surrounding pixels with similar albedos, suggesting that the spatial distribution of the material(s) responsible for the near-UV downturn is not closely tied to Callisto's geologic units.

\subsection{700 - 1000 nm Slope}\label{sec:700to1000nm}

The NIR (700 - 1000 nm) slope is mapped in Figure \ref{fig:map_700_1000nm}a. We find that much of Callisto's surface has a red 700 to 1000 nm slope, although some regions are neutral or blue. While previous analyses using Galileo/SSI imaging did not find any regions with blue NIR slopes \citep[e.g.][]{denk1999galileo}, we believe this is due to the locations of the SSI filters, which miss the downward slope starting $\sim$650 nm. Indeed, we integrated the HST spectra against the SSI green ($\sim$550 nm) and IR986 ($\sim$986 nm) bandpasses, computed the green - IR slope, and found only neutral and red slopes.  Figure \ref{fig:map_700_1000nm}b shows the moderate negative correlation between the NIR slope and bolometric albedo. Generally, the reddest material with the largest NIR slopes is dark (albedo $\lesssim$ 0.35), although not all dark material is particularly NIR red. On the leading hemisphere, the darkest material, which is concentrated to the east and south of Valhalla, has the reddest NIR slopes as well as the strongest near-UV downturn, although the most extreme values appear to be slightly offset from one another. The generally redder slopes seen on the leading hemisphere are consistent with ground based photometry and have been suggested to be related to the higher fraction of irregular satellite dust expected to be delivered \citep{morrison1974four, bottke2013black, chen2024_IrregDust}. On the trailing hemisphere, the lowest albedo pixels are not necessarily the reddest, suggesting that whatever species are responsible for the red NIR slopes must not be the only material contributing to the low albedo. The moderate correlation between the NIR red and UV red material seen on the leading hemisphere is also less apparent on the trailing hemisphere. On both hemispheres, pixels dominated by bright water-ice-rich material have blue or neutral slopes, consistent with the bluer visual color of Callisto's high-albedo material. The young impact craters, Heimdall (357\degW, 65.5\degS) and Lofn (22\degW, 56.5\degS), and their relatively extensive bright ice-rich ejecta contain pixels with the bluest measured NIR slopes.

\begin{figure}[hb!]
\centering
\includegraphics[width=\columnwidth]{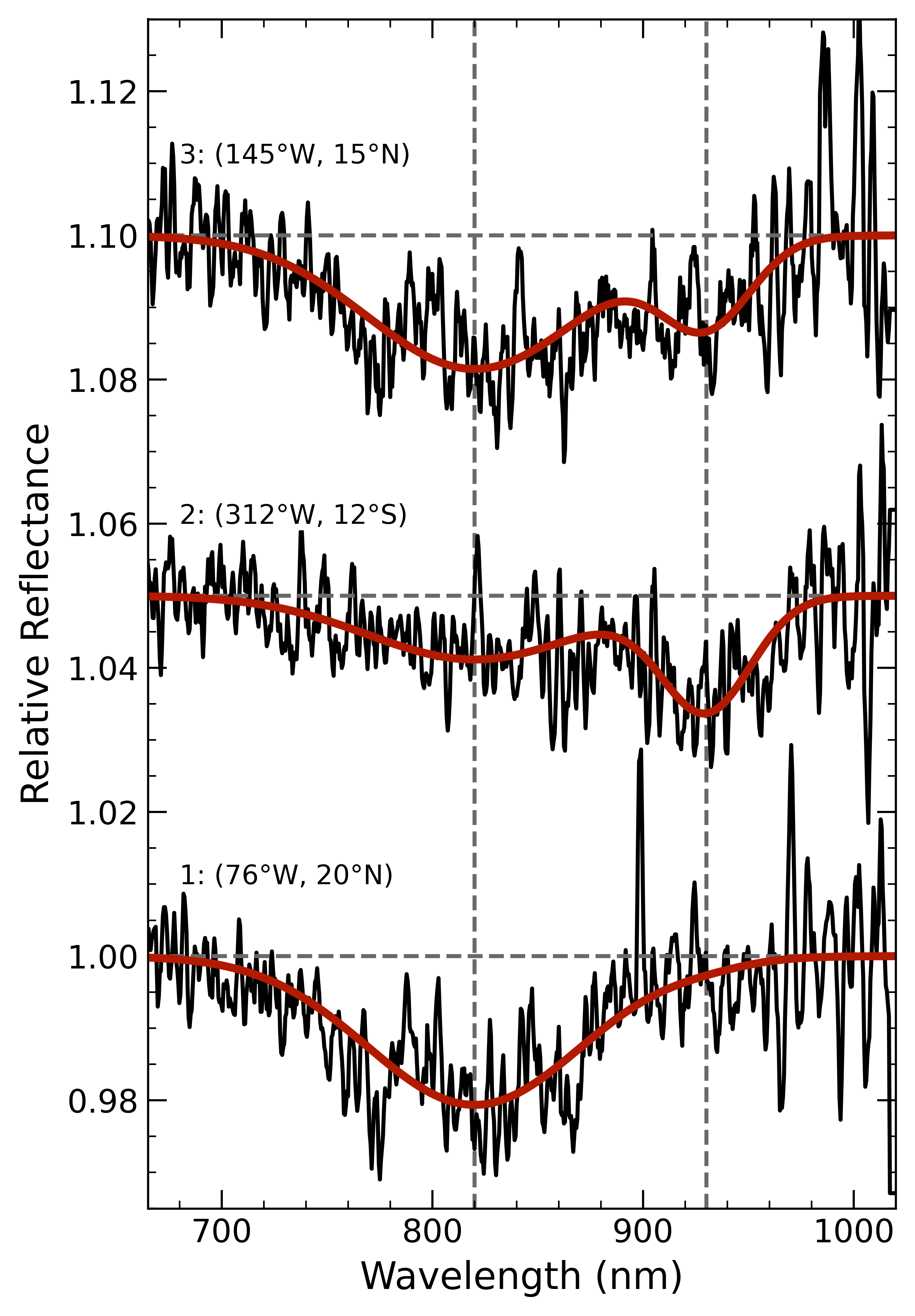}
\caption{A selection of spectra showing regions on Callisto's surface where there is a clear 820 nm absorption with little to no 930 nm absorption (1: (76\degree W, 20\degree N)), a clear 930 nm absorption with a relatively weaker 820 nm absorption (2: (312\degree W, 12\degree S)), and relatively strong 820 and 930 nm absorptions which appear to merge into a single broader feature (3: (145\degree W, 15\degree N)). The spectra have been divided by a linear continuum (horizontal dashed lines), smoothed with a 5 pixel moving average filter, and offset for clarity. The 820 nm and 930 nm band centers are indicated by vertical dashed lines and a best-fit two Gaussian model is over-plotted in red to help guide the eye. The location of each spectra is indicated on the map in Figure \ref{fig:820nm_930nm}a.}
\label{fig:820_930_diff}
\end{figure}

\begin{figure*}[p!]
\includegraphics[width=\textwidth]{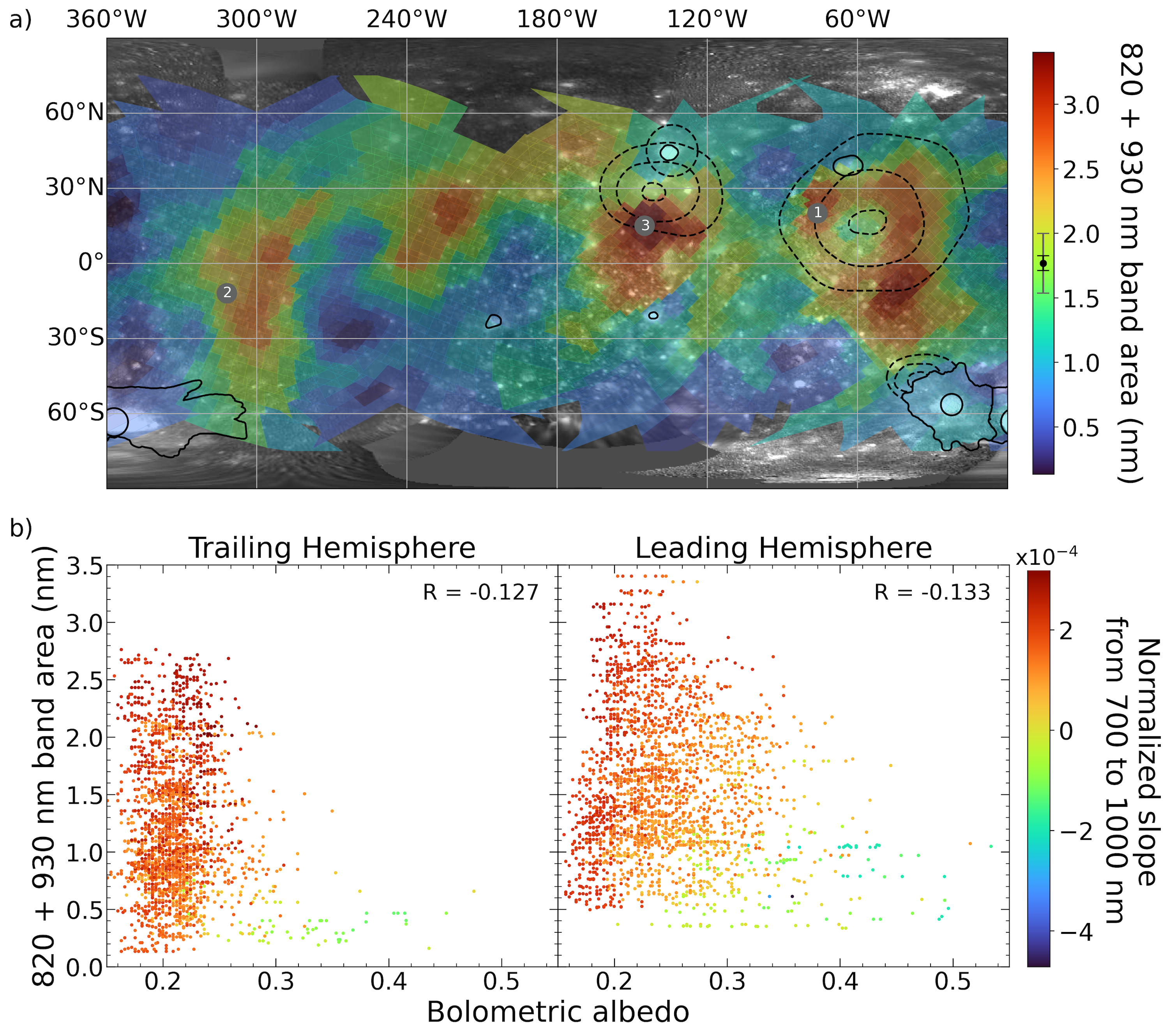}
\caption{Map showing the calculated band area of the combined 820 and 930 nm absorption features (a), along with scatter plots (b) demonstrating that there is no notable correlation between the relative albedo of the dark material and the strength of the 820 and 930 nm bands and the very weak anti-correlation can largely be explained by the tendency for the highest-albedo pixels to have little to no sign of an 820 or 930 nm feature. Pixels with the strongest 820 and 930 nm absorptions tend to have fairly red NIR slopes, especially on the trailing hemisphere where almost all of the NIR reddest pixels (700 to 1000 nm slope $\gtrapprox 2\cdot10^{-4}$) have a relatively strong combined band strength. The estimated 1$\sigma$ error for the calculated band area of each spatial pixel ranges from $\pm$0.06 to $\pm$0.23 nm and the minimum (black) and maximum (grey) estimated error is shown on the colorbar in panel (a). The grey circles on the map indicate the locations of the spectra in Figure \ref{fig:820_930_diff} which demonstrate the overlap between the 820 and 930 nm bands.}
\label{fig:820nm_930nm}
\end{figure*}

\begin{figure*}[ht!]
\includegraphics[width=\textwidth]{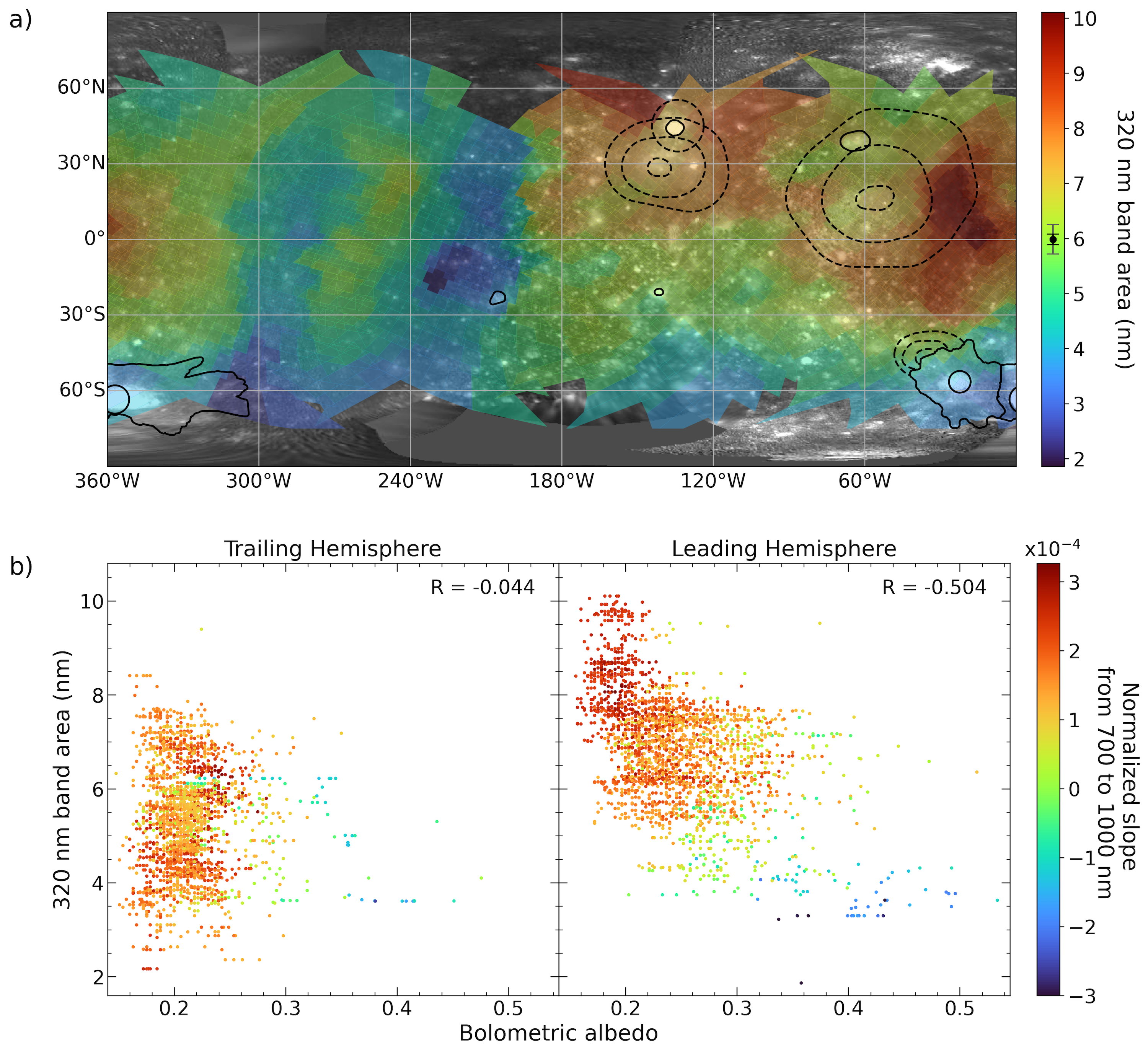}
\caption{(a) The map of the strength of the 320 nm absorption demonstrates that it is largely constrained to the regions within and surrounding the Asgard and Valhalla impact basins which leads to clear hemispheric asymmetry. The scatter plots (b) reveal a moderately strong negative correlation with albedo, with the pixels displaying a significant 320 nm absorption tending to be both dark and NIR red. The strongest measured 320 nm band areas are associated with dark and red material directly east of Valhalla, which is partially within the outer ring or trough zone. The estimated 1$\sigma$ error for the calculated band area of each spatial pixel ranges from $\pm$0.1 to $\pm$0.27 nm, and the minimum (black) and maximum (grey) estimated error is shown on the colorbar in panel (a).}
\label{fig:320nm}
\end{figure*}

\begin{figure*}[p!]
\centering
\includegraphics[width=0.9\textwidth]{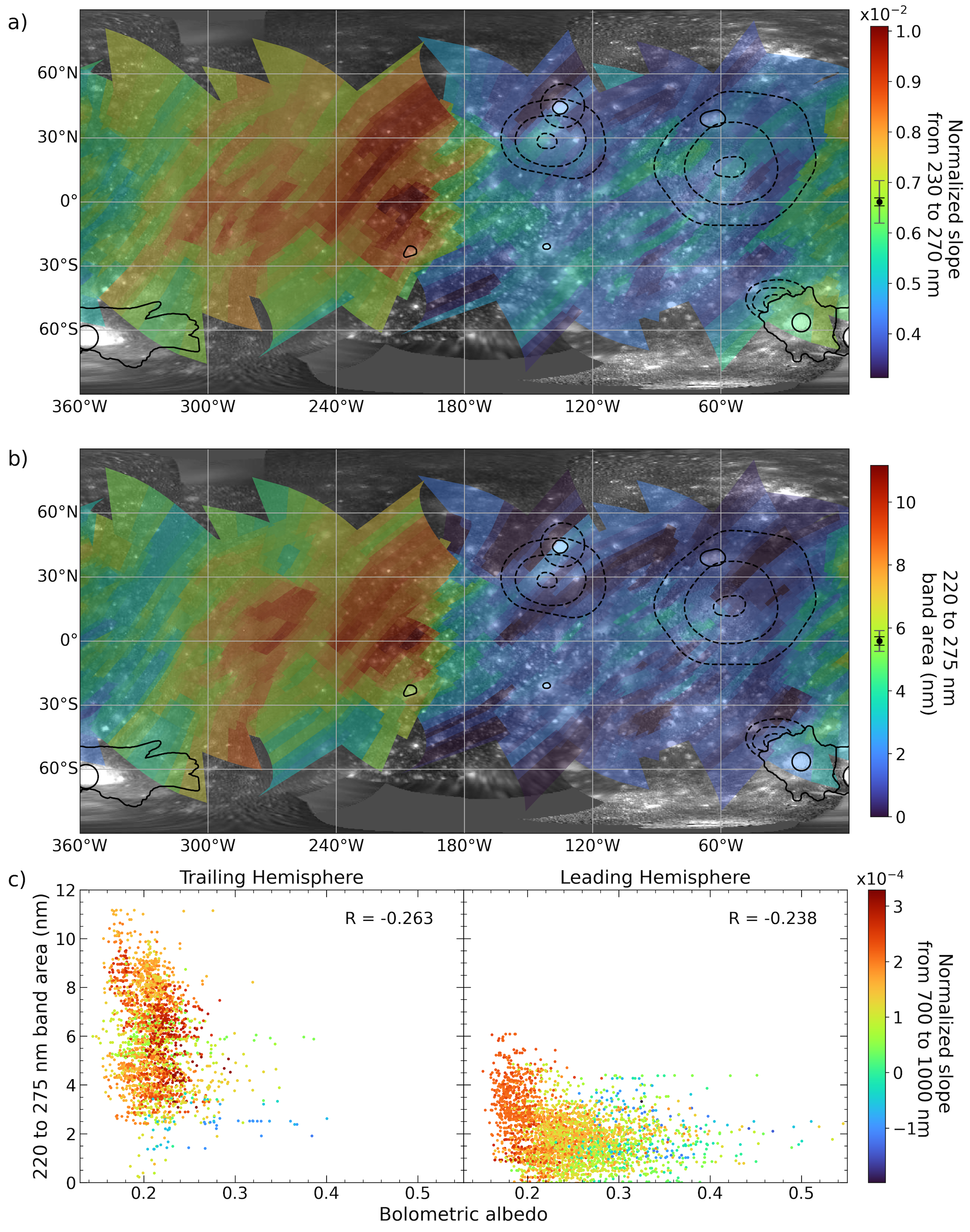}
\caption{Maps showing our two proxies for the strength of the 275 nm absorption edge- (a) the normalized slope from 230 to 270 nm , and (b) the integrated band area (from 220 nm to 275 nm) after dividing by a base spectrum constructed from five pixels identified as having the weakest 275 nm absorption edge. The feature is constrained to the trailing hemisphere and is likely associated with the stronger irradiation environment, although the somewhat irregular distribution suggests that it may require radiolysis of an unevenly distributed endogenic species. (c) Scatter plots demonstrating that the 275 nm absorption edge is uncorrelated with the bolometric albedo. The estimated 1$\sigma$ error for the calculated 230 to 270 nm slope and 220 to 275 nm band area of each spatial pixel ranges from $\pm$8.0$\times$10$^{-5}$ to $\pm$4.2$\times$10$^{-4}$ and $\pm$0.14 to $\pm$0.33 nm, respectively. The minimum (black) and maximum (grey) estimated error is shown on the colorbars in panels (a) and (b).}
\label{fig:275nm_edge}
\end{figure*}

\begin{figure*}[p!]
\centering
\includegraphics[width=0.85\textwidth]{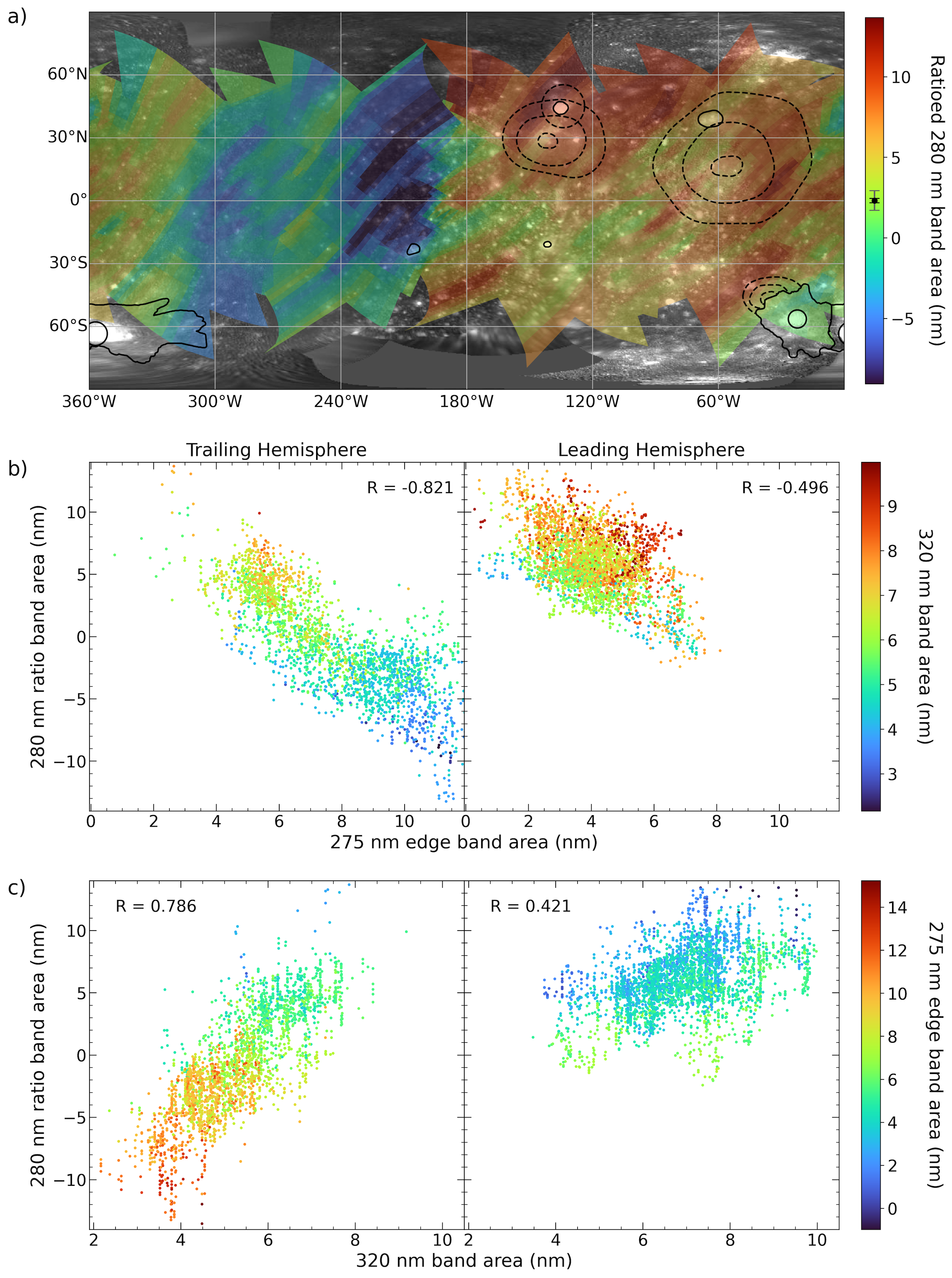}
\caption{(a) Map showing the calculated band area of the 280 nm ratio feature for spectra ratioed with the center of the trailing hemisphere. Negative values represent pixels where the ratio spectrum is above the linear continuum and there is no apparent 280 nm feature present. As expected, the band is strongest on the leading hemisphere and largely absent from the trailing hemisphere. (b, c) Scatter plots demonstrating that the 280 nm ratio feature is strongly anti-correlated with the 275 nm absorption edge, especially on the trailing hemisphere and strongly positively correlated with the 320 nm absorption band. We interpret the 280 nm ratio band to be a direct consequence of these two absorption bands, which map strongly to separate hemispheres, and find that additional absorbers on the leading hemisphere (like \ch{SO2}) are not required to explain the spatial distribution of the ratio feature. The estimated 1$\sigma$ error for the calculated band area of each spatial pixel ranges from $\pm$0.13 to $\pm$0.61 nm, and the minimum (black) and maximum (grey) estimated error is shown on the colorbar in panel (a).}
\label{fig:280nm}
\end{figure*}

The smooth bright central zones of the Utgard, Asgard and Valhalla impact basins, which have comparable near-UV downturns as the rest of the basin material, have neutral or slightly blue NIR slopes while the inner ridge and trough zones (as defined in \citet{greeley2000_GalileoViewsGeologya}) and surrounding regions are red in the NIR. As discussed in \citet{greeley2000_GalileoViewsGeologya}, after careful analysis of Galileo SSI imagery it remains uncertain whether this bright NIR neutral material is directly associated with the central plains of the ringed impact basins, or if bright material excavated from the sub-surface by recent impacts which happen to have hit near the centers of all three impact basins excavated ice-rich material that covered the otherwise NIR red surface of the large ringed structures. Based on our mapping results, however, we suggest that a direct association with the smooth central zones of the impact basins is the more likely explanation, as nearby bright impacts in the southern portion of the inner and outer ring of Valhalla, and near the eastern edge of the inner ring at Asgard do not have NIR slopes which differ significantly from the surrounding material.

\subsection{820 and 930 nm Absorption Features}\label{sec:820_930nm}
As seen in Figure \ref{fig:820_930_diff}, the relative ratio of the 820 nm and 930 nm absorptions varies across Callisto's surface, but both features are present in a majority of pixels with either band, hinting at a possible relation between the two. Because of the tendency to form what appears to be a single broad absorption, the combined strength of the 820 and 930 nm absorptions is mapped in Figure \ref{fig:820nm_930nm}a. On the leading hemisphere, dark material within the inner and outer rings and directly south of the Valhalla impact basin show relatively strong combined 820 and 930 nm band strengths (see Figure \ref{fig:820nm_930nm}a). Visual inspection of the spectra reveal that much of the Valhalla impact basin has a strong 820 nm band and the 930 nm band is present in the south east section of the inner and outer rings but much weaker in the rest of Valhalla. Indeed, the region at the north west corner (near 76\degW, 20\degN) has a strong 820 nm band with no apparent 930 nm feature, as shown in Figure \ref{fig:820_930_diff}. The relatively bright and smooth central zone and the bright material associated with the recent impactors at the north-western corner of Valhalla ($\sim$65\degW, 40\degN) are the only regions within the impact basin with a relatively weak 820 and 930 nm combined band strength. Much of the terrain surrounding the Valhalla impact basin also has a very weak 820 and 930 nm band strength, suggesting that the presence of these features may be directly related to the Valhalla impactor itself. The dark material to the south of the Asgard impact basin and within its inner and outer rings also show relatively strong 820 and 930 nm absorptions, except where the bright material from the recent impactors Doh (141.4\degW, 30.6\degN) and Burr (134.5\degW, 42.7\degN) apparently cover the dark material with water-ice rich ejecta. Two additional large patches on Callisto's trailing hemisphere also show strong 820 and 930 nm combined band strengths, while both features are absent from much of the rest of the surface.

\subsection{320 nm Absorption Feature}\label{sec:320nm}

As can be seen in Figure \ref{fig:320nm}a, Callisto's 320 nm absorption is constrained to the regions within and surrounding the Asgard and Valhalla impact basins. Pixels with a 320 nm absorption also tend to be dark and both UV and NIR red with the strongest measured band areas associated with the dark red material partially within the outer ring directly east of Valhalla. The bright material associated with fresh impact craters within Asgard and Utgard, which lack strong 820 nm and 930 nm absorption features and have relatively neutral NIR slopes, show strong 320 nm band strengths that are generally consistent with those measured in the darker material in and around the impact basins. This combined with the generally low band strengths seen for low- and mid- southern latitudes on the leading anti-Jovian hemisphere south of Asgard suggest that the strength of the 320 nm absorption feature may be more strongly associated with the Asgard and Valhalla impact basins than with the leading hemisphere dark material generally.

\subsection{230 and 450 nm Absorption Features}\label{sec:450_230nm}

The 450 nm absorption feature is inset on the broader near-UV downturn, which itself may be a composite of multiple overlapping absorptions with unidentified band parameters. As a result, the spectral continuum relative to the 450 nm absorption changes significantly across Callisto's surface. We find that while the band center of the 450 nm absorption feature appears to remain relatively consistent near 450 nm, the width, depth and shape of the continuum removed absorption feature varied significantly for different choices of polynomial degree and wavelengths included in the continuum fit and the true band parameters remain ambiguous. Attempts to map the spatial distribution of the 450 nm absorption feature produced qualitatively inconsistent results for different choices of the wavelengths included in a third order polynomial fit, and we find that we are unable to confidently map the distribution of this feature. Similarly, the 230 nm absorption feature is inset on the stronger 275 nm absorption edge. While the band center and width are more easily identified ($\sim$200 to 260 nm), we find that the increased noise in the UV observations, especially at the short-wavelength end of the spectrum, as well as a possible overlapping absorption feature short-ward of $\sim$220 nm seen in some pixels affects the continuum fit. Additionally, the 230 nm feature is very weak relative to the 275 nm absorption edge, which has an unknown effect on the concavity of the spectrum at wavelengths near 230 nm and may artificially enhance measurements of the 230 nm band strength in some pixels, hindering attempts to consistently define a spectral continuum across Callisto's surface. Taken together, this makes mapping the strength of the 230 nm absorption feature difficult and we are again unable to confidently map the spatial distribution of this feature.

While our attempts to map the strength of the 230 nm and 450 nm absorption features were unsuccessful, visual inspection reveals that both features appear to be widespread over much of the surface of Callisto. Furthermore, pixels with a strong 450 nm signature also tend to have a visually stronger 230 nm absorption while pixels with the weakest observed 450 nm absorptions show little to no sign of a 230 nm feature, hinting at a possible but unconfirmed spatial association between the two features. While the 230 nm and 450 nm absorptions appear to be relatively widespread across Callisto's surface, it is interesting to note that the features appear to be extremely weak or possibly entirely absent from pixels containing the highest albedo material. This is consistent with the absorption bands resulting from a globally distributed component of the dark material, or with a more uniform exogenic delivery that is subsequently covered up by the fresh impact ejecta typically associated with the bright water-rich material seen at the global scales considered here.

\subsection{275 nm Absorption Edge}\label{sec:275nm_edge}

Both measurements of the strength of the 275 nm absorption edge are mapped in Figure \ref{fig:275nm_edge}. Qualitatively, the two maps agree quite well and the 230 to 270 nm slope and 220 to 275 nm integrated band area are strongly correlated ($R = 0.914$). It is clear from these maps that Callisto's 275 nm absorption edge is constrained to the trailing hemisphere, suggesting that it may be associated with the fast rotating plasma bombarding Callisto's trailing hemisphere \citep{johnson2004radiationchapter}. However, its distribution does not match the bullseye pattern which is expected for species that are radiolytically produced from this plasma bombardment (e.g. \ch{CO2}) \citep{hibbitts2000_DistributionsCO2SO2, cartwright2024_RevealingCallistoCarbonrich}. The 275 nm absorption may require both the more intense irradiation environment of the trailing hemisphere and endogenic material which is not evenly distributed, explaining its absence on the leading hemisphere as well as its uneven distribution across the trailing hemisphere. The scatter plots in Figure \ref{fig:275nm_edge} c show that there is not a significant correlation between the strength of the absorption edge and the albedo. 

Both maps in Figure \ref{fig:275nm_edge} show minor diagonal striping orthogonal to the scan direction, likely due to uncorrected instrumental artifacts. However, the striping  does not correlate with surface features and has a magnitude which is smaller than the estimated 1$\sigma$ uncertainties on the band strength measurements, so we expect that this striping has a minimal effect on the interpretation of the observations.

\subsection{280 nm Ratio Feature}\label{sec:280nm}

{The strength of Callisto's 280 nm ratio feature is mapped in Figure \ref{fig:280nm}a. We expect that minor diagonal striping, as described in section \ref{sec:275nm_edge}, should have a minimal effect on the interpretation of the spatial distribution. As expected from the disk averaged observations of \citet{noll1997_DetectionSO2Callisto}, Callisto's 280 nm ratio feature is largely constrained to the leading hemisphere, although it is also present on the trailing hemisphere at longitudes larger than $\sim$300\degW. The ratio feature appears to be absent from the regions on the trailing hemisphere that show the strongest 275 nm band strengths, explaining the strong anti-correlation displayed in Figure \ref{fig:280nm}b. There is a similarly strong positive correlation with the 320 nm absorption shown in panel c. We therefore propose that the presence of the 280 nm ratio feature, and its prevalence on the leading hemisphere, may be an artifact of the strong 275 nm absorption edge present across much of the trailing hemisphere and the 320 nm absorption on the leading hemisphere. Indeed, taking the ratio of spectra containing the strongest 320 nm absorptions to those with the strongest 275 nm edge show a very strong 280 nm ratio feature with a similar band profile as that in the disk averaged ratio. See section \ref{sec:libcomp_280nm} for further discussion.

\begin{figure*}[p!]
\includegraphics[width=\textwidth]{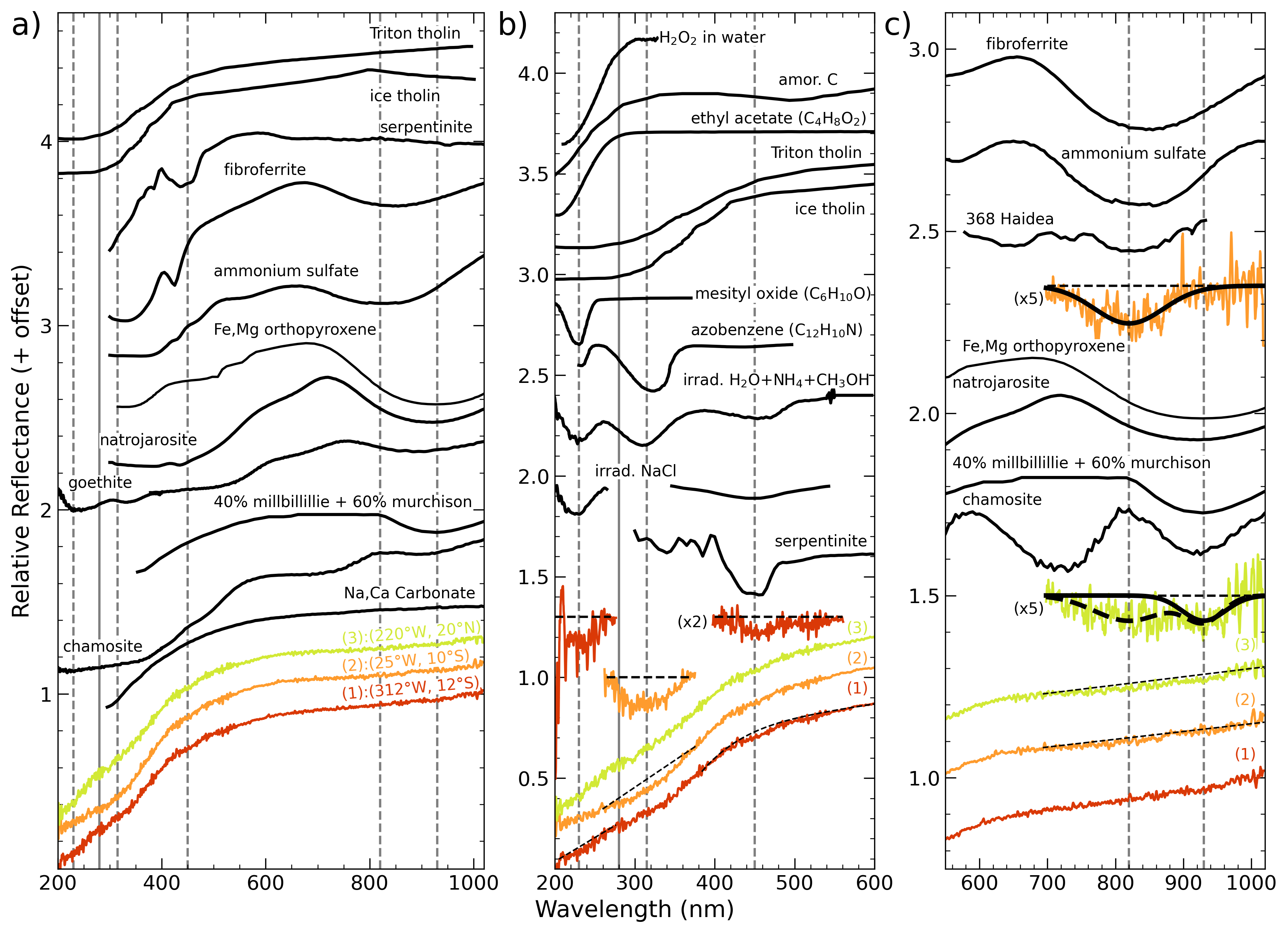}
\caption{Comparison between select Callisto spectra and some laboratory spectra discussed in the text. The color and associated number for each Callisto spectrum correspond with the locations mapped in Figure \ref{fig:roi}a and the longitude and latitude are listed in panel a. Examples of local continuum fits, indicated with black dashed lines, and corresponding continuum removed features for the 230, 320, 450, 820, and 930 nm absorption features are shown in panels b and c. The approximate positions of the band centers for the 230, 320, 450, 820, and 930 nm absorption bands are indicated with vertical dashed gray lines and the position of the long wavelength shoulder of the 275 nm absorption edge is indicated by the solid gray line. Individual laboratory spectra are shown and labeled in black. As discussed in detail in section \ref{sec:labcomp}, the identification of specific minerals to explain each UV and visible absorption feature on Callisto is difficult, and identifications typically remain ambiguous. However, we can use some of the potential spectral matches, in combination with the spatial distributions of the features, to explore possible compositions. See Sections \ref{sec:labcomp} and \ref{sec:implications} for a detailed discussion. Laboratory spectra in panel a include the Na- and Ca- bearing carbonate shortite (PSF sample CRB126), the phyllosilicate chamosite (PSF sample CHM102), an intimate mixture of the orthopyroxene rich Millbillillie eucrite meteorite and phyllosilicate rich Murchison CM2 chondrite \citep{cloutis2013spectral}, the iron oxide goethite (USGS sample WS222), the sodium sulfate natrojarosite (PSF sample SPT108), the Fe,Mg orthopyroxene enstatite (En30; \citet{klima2007spectroscopy}), ferric ammonium sulfate (PSF sample SPT104), the iron sulfate fibroferrite \citep{cloutis2006detection}, the phyllosilicate serpentinite (PSF sample SRP105), and laboratory tholins produced by irradiation of mixtures of \ch{H2O:CH3:CO2:C2H6} ices (ice tholin; \citet{mcdonald1996production}) and \ch{N2:CH4} ices (Triton tholin; \citet{cruikshank2005spectroscopic}). Panel b includes the continuum removed 450 nm feature from the serpentinite spectrum in panel a, continuum removed spectra of irradiated \ch{NaCl} at Europa-like conditions \citep{brown2022mid}, continuum removed spectra of irradiated mixtures of \ch{H2O:NH4:CH3OH} ices that are warmed to 120 K post-irradiation \citep{wong2019hubble}, the Triton and ice tholins from panel a, spectra of the liquid phase organics azobenzene, mesityl oxide, and ethyl acetate from NIST \citep{NISTWebbook}, amorphous carbon \citep{duley1984refractive}, and 0.16\% \ch{H2O2} in water ice \citep{carlson1999_HydrogenPeroxideSurface}. Panel c includes continuum removed spectra of chamosite, the Millbillillie eucrite and Murchison CM2 chondrite mixture, natrojarosite, orthopyroxene, ferric ammonium sulfate, and fibroferrite from panel a, as well as a spectrum of the asteroid 368 Haidea from \citet{vilas1994iron}.}
\label{fig:comp}
\end{figure*}

\section{Comparison of Spectral Features to Laboratory Spectra}\label{sec:labcomp}

We have completed a thorough search through the United States Geological Survey (USGS) \citep{kokaly2017_USGSSpectralLibrary}, Winnipeg Planetary Spectrophotometer Facility (PSF) \citep{cloutis2015mineral}, Keck/NASA Reflectance Experiment Laboratory (RELAB) \citep{pieters2004relab} , and Solid Spectroscopy Hosting Architecture of Databases and Expertise (SSHADE) \citep{schmitt2018solid} spectral libraries as well as a general literature search in order to find plausible spectroscopic matches to the observed Callisto spectra. In the following sections we discuss plausible matches for each of the observed features. When possible, we consider laboratory spectra obtained at cryogenic temperatures or which have been irradiated to simulate the effects of magnetospheric particle bombardment and space weathering. However, a majority of the available laboratory spectra are acquired from rock or powdered samples at room temperature and therefore do not capture the spectral changes associated with cryogenic temperatures, grain size differences, and space weathering effects, which are expected under Callisto-like conditions \citep[e.g.][]{singer1985_EffectsTemperatureRemotely, kohout2014space, escobar2018simulations}. The spectral identifications discussed in the text therefore remain preliminary and future laboratory work under conditions relevant to the Galilean satellites will be required to confirm many of these plausible spectral matches.

\subsection{820 nm and 930 nm Absorptions}
A wide variety of materials of relevance to planetary surfaces have absorption features with band centers near 820 and 930 nm, typically attributed to Fe-related electronic absorptions. We do not find any obvious or unambiguous spectral matches for either feature, however we consider a range of plausible species with interesting implications for Callisto's surface composition in the following subsections including structured silicates like orthopyroxene, sulfates and other sulfur-bearing species, phyllosilicates, and iron oxides and other weathering products. 

\subsubsection{Orthopyroxene and Iron Oxides}
Several low calcium orthopyroxenes exhibit a broad 930 nm \ch{Fe^{2+}} vibrational band \citep[e.g.][]{cloutis2002pyroxene, klima2007spectroscopy}, and iron oxides like maghemite or goethite and some iron sulfides like jarosite have a strong $\sim$930 nm \ch{Fe^{3+}} spin-forbidden absorption \citep[e.g.][]{hunt1979spectra,bishop2019_VNIR}. See Figure \ref{fig:comp} a and c for example spectra of an Fe-Mg orthopyroxene (enstatite En30 from \citet{klima2007spectroscopy}), the iron oxide goethite (USGS sample WS222), and the Na-bearing sulfate natrojarosite (PSF sample SPT108). While all three example spectra have a strong absorption band near 930 nm, it is clear that the absorption is significantly broader than what is seen on Callisto. Indeed, all laboratory spectra of typical (Ca-poor) orthopyroxenes, (hydr)oxides, and iron sulfates considered in this study have a $\sim$ 900 nm absorption that is significantly broader than is seen on Callisto, though crystal field absorptions are expected to become narrower at low temperatures. However, limited cryogenic laboratory data of enstatite (Mg-bearing orthopyroxene) \citep{singer1985_EffectsTemperatureRemotely} and hematite \citep{morris1997low}, suggest that the $\sim$900 nm absorption does not narrow significantly enough to explain the discrepancy. Additionally, iron oxides like goethite typically have an electronic Fe- absorption band near 650 nm and an Fe-OH doublet near 2.4-2.5 \um \citep{bishop2019_VNIR} which we do not see in any available spectra of Callisto, suggesting that iron oxides are unlikely to be responsible for the 930 nm absorption. 

Callisto's 820 nm absorption band is a surprisingly good match to the continuum removed spectra of 368 Haidea and 165 Loreley (see Figure \ref{fig:comp} c), two low-albedo asteroids which may have undergone less thermal alteration than the majority of C-type asteroids \citep{vilas1994iron}. \citet{vilas1994iron} suggest that this $\sim$820 nm absorption may arise from \ch{Fe^{3+}}-\ch{Fe^{2+}} charge transfer absorptions in a combination of iron alteration minerals including iron oxides like goethite and hematite. They suggest that the absorption may shift to shorter wavelengths at the lower temperatures experienced at the far edge of the asteroid belt ($\sim$195 K at 4 AU). However, we have already found that many of the iron oxides are a poor match to the 930 nm feature and the shortest wavelength band minimum for any of the iron oxides studied at 160 K in \citet{morris1985spectral} is at 860 nm, which cannot explain the 820 nm band minimum observed on Callisto. Additionally, the spectra of 368 Haidea and 165 Loreley include a prominent absorption near 650 nm which is also attributed to \ch{Fe^{3+}} and expected in goethite (see Figure \ref{fig:comp}a) and hematite, and which we do not see on Callisto. We therefore conclude that an association of Callisto's 820 nm band with iron alteration minerals, as is suggested for 368 Haidea and 165 Loreley, remains unconvincing. We note, however, that while we do not find convincing evidence of large-grained Fe-alteration minerals, we cannot rule out the presence of sub-micron or nano-phase iron or iron oxides, which are an expected space weathering product of Fe-bearing minerals \citep{clark2012surface}. Nano-phase iron and iron oxides lack the distinct absorption bands seen in their larger grained counterparts and could produce a general, but non-diagnostic, darkening and reddening of Callisto's UV and visible wavelength spectra while perhaps also weakening the strength of the mineral absorption bands \citep[e.g.][]{clark2012surface, kohout2014space, escobar2018simulations}.

\subsubsection{Sulfates}
The effects of temperature and irradiation on hydrated sulfate minerals to the visible-NIR spectrum has not been studied extensively in the lab, and it remains uncertain whether the $\sim$900 nm band common in many sulfates could narrow significantly enough to match Callisto's 930 nm feature. In addition to the strong $\sim$900 nm absorption, iron-bearing sulfates like jarosite also tend to have an Fe- related feature near 430 nm which could help explain Callisto's 450 nm band. However, the 430 nm band in Fe-bearing sulfates tends to be sharper and narrower than what is seen on Callisto, and Callisto's 450 nm band appears to be more widespread across the surface than the 930 nm absorption, suggesting that they do not arise from the same material. Sulfates also tend to have additional absorption features due to the sulfate anion (\ch{SO4^{2-}}) near 1.74-1.85 and 4-5 \um as well as an \ch{OH} related feature near 2.2-2.3 \um \citep{bishop2019_VNIR}. Galileo/NIMS observations of Callisto's dark material show possible absorption features near 2.31 and 1.8 \um \citep{moore2004callisto} which could be consistent with the presence of hydrated sulfate minerals, however NIMS spectra contain spurious radiation spikes and currently neither the 1.8 or 2.31 \um features have been confirmed with another dataset. In a recent analysis of JWST/NIRSpec observations of Callisto, \citet{cartwright2024_RevealingCallistoCarbonrich} suggest that Na-bearing sulfate could be responsible for the 3.72 \um absorption and find additional absorption features near 4.38, 4.57, and 4.92 \um which are potentially consistent with nearby absorptions in the laboratory spectrum of natrojarosite and other sulfates analyzed here. However, \citet{cartwright2024_RevealingCallistoCarbonrich} find that the 4.57 \um band is strongest at southern latitudes on the anti-Jovian hemisphere, where we find no sign of the 930 nm band, and both the 4.38 and 4.92 \um bands, while not mapped, are somewhat stronger in the average trailing hemisphere JWST spectrum than in the leading (anti-Jovian) hemisphere spectrum. It therefore remains uncertain whether Na-bearing sulfates could be responsible for Callisto's 930 nm absorption or any of the NIR features, although it is clear from their differing spatial distributions that a single sulfate mineral cannot be responsible for all of these features.

There are some \ch{Fe^{3+}}-sulfates that have a band minimum between 800 and 820 nm (e.g. rhomboclase, fibroferrite) \citep{cloutis2006detection}, however the absorptions in available laboratory spectra are broader than Callisto's 820 nm band, and as previously discussed for the 930 nm feature, it remains uncertain whether the crystal field bands in sulfate minerals will narrow sufficiently with temperature.  Interestingly, ferric ammonium sulfate has a band center at 820 nm with a narrower band width than other non-\ch{NH_4}-bearing sulfates, offering the best match to Callisto's 820 nm feature out of all of the laboratory sulfate spectra considered (see PSF sample SPT104 in Figure \ref{fig:comp} a, c). While ammonia-bearing compounds have been suggested on Callisto based on a possible weak band near 3.05 \um, spectroscopic evidence for \ch{NH_4} on Callisto is ambiguous at best \citep[e.g.][]{calvin1993spectral, cartwright2024_RevealingCallistoCarbonrich} and \ch{NH_4^+} bands near 2.55-2.65 \um have not been identified in existing NIR datasets. However, there are a surprising number of recent detections of \ch{NH_4}-bearing species on surfaces within the orbit of Saturn, including ammoniated phyllosilicates on the main-belt asteroids Ceres and Hygiea \citep{de2015ammoniated, rivkin2025_CeresJWST}, possible detections of \ch{NH_4}-bearing salts or carbonates on Ganymede from Juno/JIRAM \citep{tosi2024salts}, and the detection of ammoniated phyllosilicates on the Himalia family of Jovian irregular satellites \citep{sharkey2025_JWSTRevealsVaried}. This suggests that not only is the presence of endogenic \ch{NH_4}-bearing species on Callisto plausible, but some amount of \ch{NH4}-bearing species are almost certainly delivered to Callisto's surface by the prograde irregular satellite dust sourced from the Himalia family \citep{bottke2013black, chen2024_IrregDust}. The 820 nm band in the ferric ammonium sulfate laboratory spectra is much stronger than any of the other NIR features, and if an ammonium-bearing sulfate is responsible for Callisto's very weak 820 nm absorption, it is unlikely that any additional absorptions (e.g. the very weak 1.7-1.85 and 4-5 \um \ch{SO4^{2-}} bands) would be discernible on Callisto.

\subsubsection{\ch{Fe^{2+}}- Phyllosilicates}
Phyllosilicates might provide a better match to Callisto's 930 nm absorption as several iron-phyllosilicates common in CM chondrites show an \ch{Fe^{2+}} band near 900-940 nm with a band width that is more comparable, although still slightly broader, than Callisto's 930 nm feature \citep{cloutis2011spectralCM}. An example spectrum of chamosite (PSF sample CHM102), an Fe-rich chlorite, is shown in Figure \ref{fig:comp}a, and the polynomial continuum removed feature is compared with Callisto's 930 nm band in panel c. The band is much narrower than those seen in most orthopyroxenes, (hydr)oxides, and sulfates, providing a much closer match to Callisto, and it may narrow further at cryogenic temperatures although this has not yet been tested in the lab. The possible presence of phyllosilicates within Callisto's dark material is supported by the modeling work of \citet{calvin1991_ModelingReflectanceSpectrum} and \citet{calvin1995_SpectraIcyGalileanc}, who demonstrate that the non-ice component of Calliso's spectrum from 0.2 to 4 \um can be well-fit by models including the phyllosilicate serpentine. However, these models are based on very low resolution ground-based spectra from before Galileo arrived at the Jovian system and more recent observations with Galileo/NIMS and JWST \citep[e.g.][]{moore2004callisto, cartwright2024_RevealingCallistoCarbonrich} do not show any unambiguous sign of the $\sim$2.7 \um characteristic phyllosilicate band \citep{bishop2019_VNIR}, although it is possible that this feature may be obscured by a stronger 3 \um band associated with \ch{H2O} ice and other more abundant hydrated minerals. The laboratory spectrum of chamosite and other Fe/Mg phyllosilicates is a reasonable match to several additional NIR features on Callisto, including the OH band near 2.25-2.35 \um which is near the potential 2.31 \um band seen in some NIMS spectra. Additionally, chlorites like chamosite have a weak doublet near 3.4 and 3.5 \um which is very close to the 3.43 and 3.51 \um bands seen in NIMS observations and typically attributed to carbon-bearing minerals \citep[e.g.][]{mccord1998_NonwatericeConstituentsSurfacea, bishop2019_VNIR}. However, we note that while \citet{cartwright2024_RevealingCallistoCarbonrich} suggest that the weak 3.43 and 3.51 \um absorptions seen in the JWST/NIRSpec observations are likely real based on their correspondence to features detected in past NIMS observations, similar features detected in NIRSpec spectra of Ganymede are suggested to arise from calibration artifacts \citep{bockelee2024composition}. The bands in the JWST observations have a similar spectral contrast as other artifacts and are significantly weaker than would be expected based on the NIMS observations, which were heavily plagued by spurious radiation spikes. While it is therefore possible that the 3.43 and 3.51 \um bands are spurious and not present on Callisto, ground-based observations also show a possible weak band near 3.4 \um \citep{cartwright2020_EvidenceSulfurbearingSpecies} and the recent JWST spectra of the Jovian irregular satellites reveal a stronger 3.43 \um band attributed to aliphatic organics   \citep{sharkey2025_JWSTRevealsVaried}, perhaps suggesting that these bands are real on both Ganymede and Callisto, but related to C-H bearing organics delivered by the irregular satellite dust rather than phyllosilicates. 

Additionally, there is no evidence in any of our HST spectra for the relatively strong $\sim$700 nm \ch{Fe^{2+}-Fe^{3+}} charge transfer absorption seen in the chamosite spectrum in Figure \ref{fig:comp} that is typical in iron-bearing phyllosilicates, which are usually found in mixed valence. If single valence \ch{Fe^{2+}}-phyllosilicates are responsible for Callisto's 930 nm feature, then they are apparently uniquely devoid of \ch{Fe^{3+}} which is responsible for the $\sim$700 nm absorption attributed to phyllosilicates in many solar system bodies including carbonaceous chondrites, asteroids, on Ceres, and even in one Kuiper belt object  \citep[e.g.][]{rivkin2012fraction, seccull2018_phyllosilicateKBO, rizos2019spectral}. Recent irradiation experiments bombarding a sample from the Murchison CM2 meteorite with 1 keV \ch{H^+} and \ch{He^{+}} ions to simulate the solar wind resulted in reduction of some of the \ch{Fe^{3+}} within the phyllosilicate component to \ch{Fe^{2+}} or metallic iron \citep{laczniak2021characterizing}. It is therefore possible, although unproven, that the unique oxidation and irradiation environment on Callisto's surface could result in more efficient reduction of \ch{Fe^{3+}} in native (or delivered) phyllosilicates, perhaps reducing much of the \ch{Fe^{3+}} content to \ch{Fe^{2+}} or sequestering it into largely featureless weathering products like nanophase iron, and resulting in a 930 nm feature with an absent or very weak 700 nm absorption. 

\subsubsection{Meteoritic analogues for the 930 nm feature}

While low-calcium orthopyroxene and Fe-phyllosilicates are among the best contenders for explaining Callisto's 930 nm absorption, neither species offers a convincing match to the 930 nm band on its own. Orthopyroxenes require a more significant narrowing of the pyroxene band at low temperatures than is currently supported by laboratory data and the Fe-phyllosilicates require a very atypical single valence \ch{Fe^{2+}}-phyllosilicate or some other mechanism for suppressing the 700 nm charge transfer band. It turns out, however, that an intimate mixture of orthopyroxene and Fe-phyllosilicates may simultaneously explain both the narrowness of the 930 nm band and the absence of the $\sim$700 nm phyllosilicate absorption on Callisto. Indeed, \citet{cloutis2013spectral} analyze laboratory measurements of mixtures of fine-grained samples of the Millbillillie eucrite meteorite, whose spectrum is dominated by a broad $\sim$950 nm absorption attributed to \ch{Fe^{2+}}-rich low-calcium pyroxene, and the Murchison CM2 carbonaceous chondrite with a spectrum characterized by an overall red slope and shallow $\sim$700-1200 nm absorption feature which is a composite of the 700 nm \ch{Fe^{2+}}-\ch{Fe^{3+}} charge transfer band and the $\sim$920 nm \ch{Fe^{2+}} crystal field transition in phyllosilicates (predominantly cronstedtite and serpentine). They find that for intimate mixtures of the two meteorites, the resulting spectrum is dominated by the $\sim$900 nm pyroxene band and there is no compelling evidence for the presence of the expected phyllosilicate absorption band near 700 nm even for mixtures with 60 wt.\% phyllosilicate rich CM2. Additionally, they show that increasing amounts of the murchison CM2 in the intimate mixtures shifts the pyroxene band to shorter wavelengths (933 nm for 60 wt.\% CM2) and results in a shallower band depth with a small but measurable reduction in the band width. A mixture of 40\% orthopyroxene rich Millbillillie with 60\% phyllosilicate rich Murchison from \citet{cloutis2013spectral} is shown in Figure \ref{fig:comp}c and is remarkably consistent with the weak and narrow 930 nm band measured on Callisto.

If Callisto's 930 nm band arises from a mixture of orthopyroxene and phyllosilicates, it may not be surprising that it is apparently associated with the large impact basins. Typical silicates like orthopyroxene are common throughout the solar system and may be present in Callisto's endogenic rocky material or be delivered from the giant impactors themselves. Heat generated by the basin forming impacts into Callisto's ice-rich surface could have produced melt which resulted in the aqueous alteration of some of the original orthopyroxene into phyllosilicates.

\subsection{230 nm and 450 nm Absorptions}
We were unable to map the spatial distributions of the weak 230 nm and 450 nm absorption features (see Section \ref{sec:450_230nm}). However, visual inspection of the individual spectra reveals that both features are present across a majority of Callisto's surface, and their strengths may be correlated. If the 230 nm and 450 nm absorptions arise from the same material, they may be a reasonable spectroscopic match to irradiated \ch{NaCl}, which was recently detected within chaos terrains on Europa's leading hemisphere via color center absorptions at 230 nm and 450 nm \citep{trumbo2019sodium, trumbo2022new}. Figure \ref{fig:comp}b shows a continuum removed laboratory spectrum of irradiated \ch{NaCl} at Europa-like conditions \citep{brown2022mid} along with the continuum removed 230 and 450 nm absorptions on Callisto. The 450 nm band center is a reasonable match to the \ch{NaCl} laboratory spectrum as well as Europa's 450 nm feature, but Callisto's continuum removed feature appears to be significantly narrower than expected for \ch{NaCl}. However, the 450 nm absorption feature is inset on the broader near-UV downturn and $\sim$415 nm slope break, and it is difficult to discern the spectral continuum in this region of Callisto's spectrum, so the exact band shape and width of Callisto's 450 nm feature remains ambiguous. Indeed, as discussed in Section \ref{sec:450_230nm}, changes to the wavelengths included in the continuum polynomial fit significantly alter the apparent 450 nm band width and shape and we conservatively choose the continuum fit displayed in Figure \ref{fig:comp}b to highlight the deepest, most visually obvious portion of the absorption. We suspect that the true band width of Callisto's 450 nm absorption may be broader than the narrow region captured by this choice of continuum, perhaps extending to $\sim$370 nm on the short wavelength end where there is a small divet in many of the spectra with strong 450 nm absorption's that is visually similar to that seen for the 450 nm \ch{NaCl} feature on Europa. The 230 nm absorption observed on Callisto is also a reasonable match, in both band center and width, to the \ch{NaCl} laboratory spectra as well as Europa's 230 nm absorption.

Alternatively, Callisto's 450 nm absorption is a reasonable visual match to the 450 nm spin-forbidden electronic transition of tetrahedrally-coordinated \ch{Fe^{3+}} in hydrated minerals like serpentines \citep{burns1993mineralogical} (see serpentinite spectrum from PSF sample SRP105 in Figure \ref{fig:comp}). Indeed, models including serpentine have been shown to fit Callisto's 0.2 to 4 \um spectrum reasonably well \citep{calvin1991_ModelingReflectanceSpectrum, calvin1995_SpectraIcyGalileanc}, albeit at relatively low spectral resolution, and many serpentines have a characteristic OH band near 2.3 \um which may explain the potential 2.31 \um band seen in some NIMS spectra. If this feature is related to a phyllosilicate like serpentine, the apparently widespread distribution of the 450 nm band suggests that it is either delivered from irregular satellite dust or excavated from the subsurface, and it cannot be directly related to the potentially phyllosilicate-bearing material responsible for the geographically constrained 930 nm absorption. 

\subsection{275 nm Absorption Edge}
The 275 nm absorption edge, seen predominantly on Callisto's trailing hemisphere, has been attributed to the presence of hydrogen peroxide (\ch{H2O2}) produced via irradiation of water ice \citep{hendrix1999_HYDROGENPEROXIDEICY}. The laboratory spectrum of 0.16\% \ch{H2O2} in water ice \citep{carlson1999_HydrogenPeroxideSurface}, as shown in Figure \ref{fig:comp}b, is indeed a reasonable visual match to Callisto's 275 nm absorption edge. However, neither Galileo/NIMS or the more recent JWST/NIRSpec observations of Callisto show any evidence of a 3.5 \um peroxide absorption which is seen on Europa and Ganymede \citep{carlson1999_HydrogenPeroxideSurface, trumbo2019_H2O2ChaosTerrain, trumbo2023_HydrogenPeroxideGanymede}. It is possible that the concentration of \ch{H2O2} on Callisto may be low enough that the strong electronic transition band near 210 nm is observed, while the much weaker 3.5 \um O-H stretch vibrational overtone is masked. Alternatively, \citet{hendrix2008_CallistoNewInsights} also suggest that the reddening of the spectral slope towards shorter UV wavelengths may be dominated by carbon-based species. In either case, the feature is strongly geographically constrained to the trailing hemisphere and we expect that the species responsible for the 275 nm absorption edge is an irradiation product associated with the stronger magnetospheric bombardment experienced by Callisto's trailing hemisphere. If the 275 nm absorption edge is associated with \ch{H2O2} in irradiated water ice, then we might expect the strongest band strengths to be associated with the bright material. However, as can be seen in Figure \ref{fig:275nm_edge}c, we find the strongest band strengths are associated with very low albedo ($\lesssim$0.2) water-poor material. This suggests that an association with organics is the more likely explanation and could indicate that the 275 nm absorption edge is caused by irradiation-induced alteration of dark carbon-rich material through some process (e.g. carbonization) which requires the higher irradiation fluxes experienced on Callisto's trailing hemisphere. Indeed, Callisto's \ch{CO2} signature is strongest on the trailing hemisphere and appears to be related to the magnetospheric bombardment, perhaps consistent with an irradiated carbon-bearing material producing both \ch{CO2} and heavily carbonized organic molecules responsible for the 275 nm absorption edge \citep{hibbitts2000_DistributionsCO2SO2,hibbitts2002co2, cartwright2024_RevealingCallistoCarbonrich}. See section \ref{sec:labcomp_320nm} for further discussion of the 275 nm absorption edge in the context of organic species on Callisto.

\subsection{320 nm Absorption and Other Plausible Organic Features}\label{sec:labcomp_320nm}

The newly identified 320 nm absorption feature, seen predominantly in and around the Valhalla and Asgard impact basins on Callisto's leading hemisphere, may be related to the 280 nm rollover seen in the Galileo/UVS spectra (210 - 320 nm), which \citet{hendrix2008_CallistoNewInsights} attribute to organic absorption. They suggest this feature is strongest in the south polar region, where they measured steep blue slopes from 280–320 nm, and propose that organics may be progressively weathered at low latitudes by UV, charged particle, and micrometeorite bombardment. Our HST/STIS observations do not extend to the south pole, but we find that the 320 nm absorption is generally weak at mid-high southern latitudes, and regions with the strongest absorption do not consistently exhibit blue 280–320 nm slopes. While the 320 nm feature could still be linked to the material responsible for the 280 nm rollover, our broader wavelength coverage reveals a spatial distribution that differs significantly from that inferred in \citet{hendrix2008_CallistoNewInsights}. Nevertheless, the 320 nm absorption remains consistent with the presence of organic molecules, long considered a plausible component of Callisto’s dark material \citep[e.g.][]{mccord1997_OrganicsOtherMolecules, mccord1998_NonwatericeConstituentsSurfacea, hendrix2008_CallistoNewInsights}, though its precise origin warrants further study.

Unsaturated organics, molecules that contain at least one double or triple carbon bond, have strong UV absorption bands associated with a \ch{$\pi$-$\pi$^*} electronic transition that often occur at wavelengths between 200 and 500 nm \citep{kalsi2007spectroscopy}, many of which could present a plausible match to Callisto's 320 nm band, 275 nm absorption edge, and possibly even the 230 and 450 nm features. However, the identification of specific organic molecules or sub-types are typically not possible from UV and visible absorption features as small changes to the molecular structure such as the inclusion of one additional double bond can shift the band position by up to 30 nm at a time \citep{kalsi2007spectroscopy}. Existing literature exploring the production (or destruction) of organics by irradiation of carbon and nitrogen bearing species, especially at Callisto-like conditions, is sparse and the influence of these organics on the UV and visible wavelength reflectance spectrum has not been well studied in the laboratory. Recent experiments simulating the effect of solar wind on the murchison CM2 meteorite showed that \ch{H^+}- and \ch{He^+}- irradiation may affect the organic functional group chemistry in different ways, with \ch{H^+} breaking down macromolecular material and increasing the abundance of low-molecular-weight free organic species while \ch{He^+} decreases the overall organic content \citep{laczniak2021characterizing}. The radiolytically induced organic chemistry on carbon-rich surfaces apparently depends on the bombarding species and energy, and the balance of constructive and destructive processes in the much more intense irradiation environment at Callisto is unknown. Further complicating the picture, laboratory studies of meteorites reveal a large diversity of organic molecules which range from simple ones like amino acids and carboxylic acids to more complex species like polycyclic aromatic hydrocarbons and macromolecular organic matter \citep[e.g.][]{cloutis2011spectralCI, cloutis2011spectralCM} and it is likely that the organic component on Callisto has a similar diversity of species resulting in a complex mixture of bands presented in the UV spectrum. With this in mind, we discuss several types of organic species below with example spectra shown in Figure \ref{fig:comp} that provide a plausible visual match to some of Callisto's UV spectral features, but we by no means suggest that any of these particular species are present on Callisto. Instead, we ask the reader to view this as a demonstration that organic molecules of a variety of types can offer a plausible match to many of Callisto's UV features. 

Recent observations of several of the Jovian irregular satellites with JWST/NIRSpec revealed strong ($\sim$1-3\% band depths) 3.4 \um absorptions attributed to aliphatic organics on all five of the observed retrograde satellites and on at least one of the prograde Himalia family moons \citep{sharkey2025_JWSTRevealsVaried}. These organics are delivered to Callisto via irregular satellite dust grains and likely explain the 3.43 \um band reported in both the leading and trailing hemispherically averaged spectrum from JWST (assuming it is not an artifact as in \citet{bockelee2024composition}) \citep{cartwright2024_RevealingCallistoCarbonrich}. The relatively weak ($\lesssim$1\%) 3.43 \um band depths measured in the Callisto JWST spectra could arise from this dust mixing with other species on the surface, or it might indicate that some of these delivered organics are radiolytically altered and incorporated into new species \citep{mccord1997_OrganicsOtherMolecules, cartwright2024_RevealingCallistoCarbonrich}. Callisto's 4.57 \um band has also been suggested to arise from C-N (or C-S) bearing organics delivered directly from irregular satellite dust \citep{cartwright2020_EvidenceSulfurbearingSpecies, cartwright2024_RevealingCallistoCarbonrich}, however the 4.57 \um band and any spectral indication of S-bearing species is notably absent from the irregular satellite JWST spectra \citep{sharkey2025_JWSTRevealsVaried}. Instead, the 4.57 \um band may arise from the irradiation of \ch{NH4}-bearing compounds delivered by the dust which mixes with native \ch{H2O} ice and carbon-rich material or \ch{CO2}, perhaps following a similar radiolytic pathway as mixtures of \ch{N-}, \ch{C-}, \ch{H-}, and \ch{O-} bearing ices which form a variety of new radiolytic products such as \ch{OCN^-} and long-chain refractory organics with C-H and C-N stretching modes \citep{accolla2018combined, cartwright2024_RevealingCallistoCarbonrich, sharkey2025_JWSTRevealsVaried}. An understanding of the spatial distribution of the 3.43 \um band and how it relates to the 4.57 \um feature on Callisto may help determine whether the organics delivered by the irregular satellite dust are cannibalized and incorporated into the more refractory organics tentatively associated with the 4.57 \um band. Regardless, irradiation of carbon-bearing and ice-rich materials have been shown to create a variety of complex organic molecules in laboratory simulations \citep[see e.g.][]{palumbo2004infrared, de2017photo, accolla2018combined, faure2021radiolytic, urso2020irradiation} and we expect that radiolytic processing plays a crucial role in the formation of organics on Callisto and is required to explain the apparently vastly different spatial distributions of the 320 nm, 3.43 \microns, and 4.57 \um absorptions if the interpretation that they result from organic materials is correct.

Some laboratory tholins, like those suggested to explain the prominent 4.57 \um absorption can also exhibit strong UV bands \citep{mccord1997_OrganicsOtherMolecules, cartwright2024_RevealingCallistoCarbonrich, gavilan2017organic}. Figure \ref{fig:comp} a and b shows the reflectance spectrum, calculated via Hapke modeling (as in \citet{hendrix2010ultraviolet}) from laboratory optical constants for an ice tholin and Triton tholin produced via irradiation of a mixture of \ch{H2O:CH3OH:CO2:C2H6} \citep{mcdonald1996production} and \ch{N2:CH4} \citep{cruikshank2005spectroscopic} ices, respectively. The start of a strong UV absorption edge in the ice tholin aligns well with the long wavelength edge of Callisto's 320 nm absorption, and the Triton tholin shows a drop-off towards the UV starting at slightly longer wavelengths. While the laboratory tholin absorption edges continue to shorter wavelengths and are therefore not a spectral match to Callisto's 320 nm feature, we can not rule out a possible contribution from such tholin like organics to the 320 nm band, or to the generally red UV slopes and near-UV downturn which are ubiquitous across Callisto's surface. Regardless, tholin-like organics are unlikely to fully explain the band width and shape of the observed 320 nm band, and additional species are required.

Interestingly, irradiated mixtures of \ch{H2O:NH4:CH3OH} ices that are warmed to 120 K post-irradiation in a recent experiment designed to explore the irradiation and heating history of the Jupiter Trojans \citep{mahjoub2016electron, poston2018visible} provide a remarkably good match to the 230, 320, and 450 nm absorption bands seen on Callisto. The continuum removed spectrum of the irradiation product after heating, as shown in \citet{wong2019hubble}, is plotted in Figure \ref{fig:comp}b. The specific causes of these UV absorptions are not identified in the laboratory analysis of \citet{poston2018visible}, however \citet{mahjoub2016electron} identify the presence of several \ch{HCOOH} and \ch{HCONH2} -bearing irradiation products in the infrared spectrum from the same set of experiments, suggesting that a variety of CH- and CN- bearing organics are formed which may explain the UV bands. While direct irradiation of \ch{NH4} and \ch{CH3OH} ices is not relevant at Callisto, perhaps the irradiation of mixtures of water-ice, ammonium-bearing species, and carbonates or \ch{CO2} can produce similar CH- and CN- bearing organics, possibly explaining the stark similarities between the 230, 320, and 450 nm absorptions seen in the laboratory irradiation product and on Callisto. While N-bearing species have not been unambiguously detected on Callisto, recent JWST observations have detected ammoniated phyllosilicates on the prograde Jovian irregular satellites Himalia and Elara which are expected to source dust, likely including \ch{NH4}-bearing compounds, that is delivered to Callisto's surface \citep{bottke2013black, sharkey2025_JWSTRevealsVaried}.
 
Several organic molecules with absorption bands near 230 nm and 320 nm, and with a shoulder near the 275 nm edge are included in Figure \ref{fig:comp} to show that organic molecules are capable of producing UV bands similar to those seen on Callisto. As can be seen in Figure \ref{fig:comp}b, the N-bearing aromatic azobenzene (\ch{Ph-N=N-Ph}) has a strong $\sim$320 nm absorption \citep{NISTWebbook} and the aliphatic ketone mesityl oxide (\ch{ CH3C(O)CH=C(CH3)2}) displays a strong $\sim$230 nm band \citep{NISTWebbook}. N-bearing aromatics like azobenzene also tend to display a second absorption, in this case near 200 nm, which could help explain Callisto's 230 nm feature. However, we do not find any clear evidence of a strong spatial correlation between Callisto's 230 and 320 nm absorption bands, suggesting a single organic sub-type or formation pathway cannot fully explain both absorptions. Additionally, the spectrum of the relatively simple carboxylate ester ethyl acetate (\ch{CH_{3}CO_{2}CH_{2}CH_{3}}) and amorphous carbon, a common product of irradiation induced carbonization of organics \citep[e.g.][]{laczniak2021characterizing}, are shown in Figure \ref{fig:comp}. Both have a strong absorption band near 200 nm which is a reasonable visual match to the 275 nm absorption edge seen predominantly on Callisto's trailing hemisphere. Amorphous carbon and other aromatic carbon rings is perhaps the simplest explanation for the 275 nm absorption edge, as we might expect the more heavily irradiated trailing hemisphere to efficiently break down (or carbonize) existing organics. This is consistent with the absence of the 320 nm band on the trailing hemisphere as well as the much weaker 4.57 \um band depth reported in \citet{cartwright2024_RevealingCallistoCarbonrich}.

The association of Callisto's UV spectral features with organic molecules remains tentative. We have completed a thorough search through the USGS, PSF, RELAB, and SSHADE spectral libraries as well as a general literature search and have not found any other plausible spectral matches for Callisto's 320 nm absorption, perhaps strengthening the case for an organic origin. The 230 nm and 450 nm absorption bands, on the other hand, are consistent with other plausible species including \ch{NaCl} and Fe-bearing silicates like serpentinite, and we do not find any strong spectroscopic evidence to prefer one explanation over the others. 
 
\subsection{280 nm Ratio Feature}\label{sec:libcomp_280nm}
\begin{figure*}[ht!]
\includegraphics[width=\textwidth]{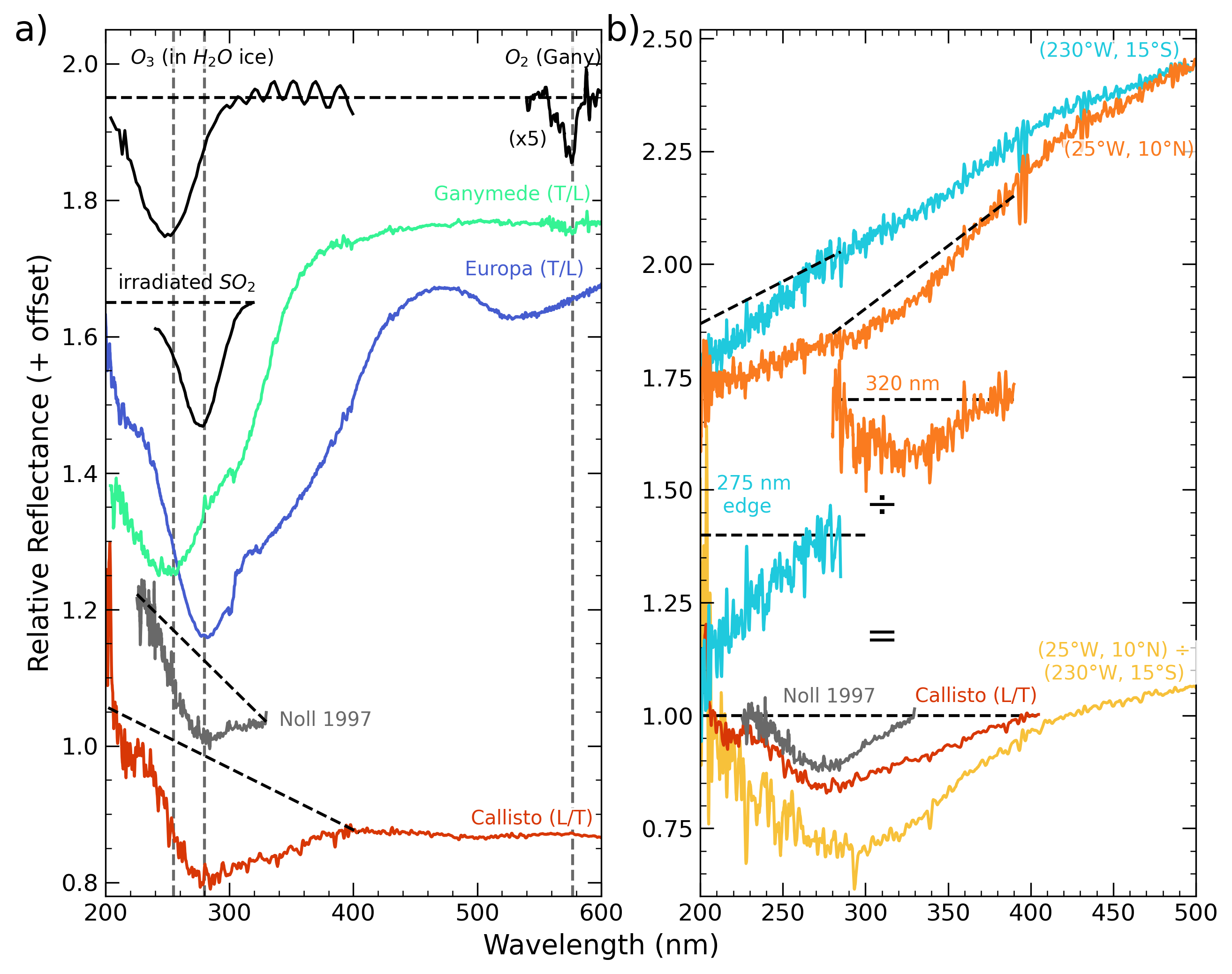}
\caption{(a) Callisto's 280 nm absorption shows up in leading hemisphere spectra when ratioed against the trailing hemisphere. Europa's trailing/leading hemisphere ratio spectrum (blue) shows a clear 280 nm band which matches the laboratory spectrum of irradiated \ch{SO2} ice that is warmed to 100 K \citep{ramachandran2023_UltravioletSpectrumReveals} and is typically attributed to \ch{SO2} created from Iogenic sulfur implantation \citep{lane1981evidence, hendrix2011europa, becker2022_MidultravioletHubbleObservations}. Ganymede's trailing/leading hemisphere ratio spectrum (teal) shows a strong band centered at 260 nm attributed to radiolytically produced \ch{O3} \citep{noll1996detection, boduch2016ion}, which also clearly matches the laboratory spectrum of \ch{O3} in water ice from \citet{teolis2006ozone}. The hemispherically averaged leading/trailing spectrum of Callisto (red) is compared with the observations of \citet{noll1997_DetectionSO2Callisto} which originally attributed this band to \ch{SO2} on the leading hemisphere. Our HST observations clearly show that the band is broader than assumed by \citet{noll1997_DetectionSO2Callisto} and is not necessarily an obvious match to \ch{SO2} or a combination of \ch{SO2}, \ch{O3}, and an unidentified band past 300 nm as suggested by \citet{ramachandran2023_UltravioletSpectrumReveals}. Instead, as demonstrated in (b) the ratio spectrum can be explained by diving a spectrum with a strong 320 nm feature, which maps predominantly to the leading hemisphere, by a spectrum with a strong 275 nm absorption edge that is constrained to the trailing hemisphere. The spectra of the two pixels which produce the deepest 280 nm ratio feature are plotted in blue and orange and their ratio (orange/blue) is shown in yellow, clearly demonstrating that the ratio of the 320 nm band/275 nm edge produces an apparent 280 nm absorption. This means that an additional absorber, like \ch{SO2} or \ch{O3} are not required to explain the 280 nm ratio feature and we do not find strong evidence for \ch{SO2} on Callisto's leading hemisphere.
}\label{fig:comp_280nm}
\end{figure*}

The 280 nm ratio feature has been attributed to the presence of \ch{SO2} \citep{lane1981evidence, noll1997_DetectionSO2Callisto} or a combination of \ch{SO2}, \ch{O3}, and an additional unidentified $\sim$300 nm band \citep{ramachandran2023_UltravioletSpectrumReveals} on Callisto's leading hemisphere. Figure \ref{fig:comp_280nm}a shows the disk-integrated ratio spectrum of Callisto's leading to trailing hemisphere alongside the more limited wavelength coverage observation from \citet{noll1997_DetectionSO2Callisto} and similarly constructed trailing to leading hemisphere ratios from HST/STIS observations of Europa and Ganymede. For Ganymede, the ratio spectrum reveals a relatively strong band near 255 nm which is attributed to radiolytically produced \ch{O3} \citep{noll1996detection, boduch2016ion} and is indeed a reasonable visual match to the laboratory spectrum of \ch{O3} in water ice \citep{teolis2006ozone}. The Europa ratio spectrum shows a strong band centered near 280 nm, which is attributed to the radiolytic production of \ch{SO2} from sulfur bombardment of the trailing hemisphere \citep{lane1981evidence, hendrix2011europa, becker2022_MidultravioletHubbleObservations} and is indeed a fair visual match to the laboratory spectrum of irradiated \ch{SO2} ice that is subsequently warmed to 100 K \citep{ramachandran2023_UltravioletSpectrumReveals} shown in Figure \ref{fig:comp_280nm}a. While both Europa and Ganymede appear to have additional contributions to the broader UV features in their hemispherical ratio spectra, they both show a clear inset band, which is consistent with the band center and width of \ch{O3} on Ganymede and \ch{SO2} on Europa. \citet{ramachandran2023_UltravioletSpectrumReveals} suggest that the hemispherically averaged leading to trailing ratio spectrum for Callisto, as obtained by \citet{noll1997_DetectionSO2Callisto}, can be well fit by a combination of three Gaussian's with band parameters matching \ch{SO2}, \ch{O3}, and an additional unknown absorber with a band center long-wards of 300 nm. However, as can be seen in Figure \ref{fig:comp_280nm}a, as well as the continuum removed feature shown in panel b, our HST/STIS observations reveal that the width of the 280 nm ratio feature is significantly wider than assumed from the limited wavelength coverage observations from \citet{noll1997_DetectionSO2Callisto}, and additional components on both the short and long wavelength edges would likely be required to fit this broader feature. 

Laboratory experiments have shown that, at the temperatures of the Jovian satellites, \ch{SO2} and \ch{O3} will efficiently react making detection of these molecules in the same vicinity unlikely \citep{loeffler2016eating}, and suggesting that if these species are responsible for the ratio feature, then the relative strengths of the \ch{SO2} and \ch{O3} absorption bands should vary across Callisto's surface. Yet, we find no evidence for any distinct absorption bands near either the \ch{O3} or \ch{SO2} band centers in Callisto's hemispherical ratio spectrum or in ratio spectra for any of the individual spatial pixels. We therefore suggest that Callisto's 280 nm ratio feature is unlikely to be caused by an abundance of \ch{SO2} and \ch{O3} on the leading hemisphere. 

Instead, we suggest a simpler explanation for Callisto's 280 nm ratio feature that only involves the 320 nm absorption band and 275 nm absorption edge, which are clearly seen in individual spatial pixels and mapped in Figures \ref{fig:320nm} and \ref{fig:275nm_edge}. Spectra of the two spatial pixels that produce the largest 280 nm ratio feature ((25\degW, 10\degN) and (230\degW, 15\degS)), identified by iteratively dividing every combination of leading and trailing hemisphere pixel pairs and calculating the residual band area, are shown in Figure \ref{fig:comp_280nm}b along with their ratio spectrum. It is clear that the leading hemisphere pixel has a strong 320 nm absorption band and the trailing hemisphere spectrum has a strong 275 nm absorption edge. The continuum removed bands are shown below the spectra illustrating that the division of a spectrum containing a clear 320 nm absorption, which maps predominantly to the leading hemisphere (Figure \ref{fig:320nm}a) by a spectrum containing a strong 275 nm absorption edge, seen predominantly on the trailing hemisphere (Figure \ref{fig:275nm_edge}a), can reproduce the broad feature with a 280 nm band minimum seen in Callisto's ratio spectra without requiring any additional absorptions from species like \ch{SO2} or \ch{O3}. This is consistent with our mapped spatial distribution for the 280 nm ratio feature (Figure \ref{fig:280nm}a) and, as expected, a clear anti-correlation with the strength of the 275 nm absorption edge and positive correlation with the 320 nm band area are shown in Figure \ref{fig:280nm} b and c. We do note, however, that while the 280 nm band is better explained by the 275 nm edge and 320 nm absorption, we cannot rule out a small additional contribution from either \ch{SO2} or \ch{O3}.

\subsection{Near-UV Downturn}\label{sec:labcomp_nearUV}

A similar near-UV downturn is seen on Callisto and the other Galilean satellites---Io, Europa, and Ganymede, as well as on the Saturnian satellites and on small bodies throughout the solar system \citep{spencer1995charge, hendrix2018icy, trumbo2020endogenic, trumbo2022_SpectroscopicMappingIo}. This downturn has been attributed, at least in part, to the presence of sulfur allotropes for Io and Europa's trailing hemisphere, which is heavily bombarded by Iogenic sulfur ions \citep[e.g.][]{spencer1995charge, trumbo2020endogenic, trumbo2022_SpectroscopicMappingIo}. Some amount of Iogenic sulfur is likely delivered to Callisto via Jovian magnetospheric bombardment \citep{cooper2001energetic, liuzzo2019energetic} and \citet{cartwright2020_EvidenceSulfurbearingSpecies} suggest that an absorption near 4.02 \um seen primarily on the leading hemisphere could plausibly be explained by the presence of thermally-altered sulfur allotropes from Io or from irradiation of \ch{H2S}-rich dust grains originating from Jupiter's retrograde irregular satellites. While it is plausible that sulfur-bearing species could contribute to Callisto's near-UV downturn, we find that the presence of \ch{SO2} on Callisto's leading hemisphere is not required to explain the observed 280 nm ratio feature, suggesting that \ch{SO2} and perhaps other sulfur-bearing species may be less abundant on Callisto than previously thought. This is consistent with the more recent JWST/NIRSpec results from \citet{cartwright2024_RevealingCallistoCarbonrich} who favor the hypothesis that Callisto's spectral properties are dominated by carbon, with sulfur perhaps providing only minor contributions. Additionally, recent observations of the Jovian irregular satellites with JWST/NIRSpec show no reliable evidence for S-bearing species \citep{sharkey2025_JWSTRevealsVaried}, likely ruling out the irregular satellites as a significant source of sulfur on Callisto. Iogenic sulfur within the Jovian magnetospheric plasma is expected to preferentially bombard Callisto's trailing hemisphere \citep{cooper2001energetic, moore2004callisto}, although recent simulations accounting for interactions between the plasma and Callisto's induced magnetic field and ionosphere suggest that the trailing hemisphere may sometimes be partially shielded from energetic ions, which are instead focused towards the sub-Jovian (0\degW, 0\degN) and anti-Jovian (180\degW, 0\degN) apexes \citep{liuzzo2019energetic}. The observed spatial distribution of the near-UV downturn (see Figure \ref{fig:uvdropoff}b) does not show any particular enhancement associated with the trailing hemisphere or near the sub- and anti- Jovian apexes, suggesting that magnetospheric plasma is not likely to supply the primary material(s) responsible for the downturn. When combined with the absence of S-bearing species in the delivered irregular satellite dust, this likely rules out sulfur allotropes as a significant contributor to Callisto's near-UV downturn.

\citet{trumbo2020endogenic} has also suggested that irradiation damaged water ice may contribute to the near-UV downturn on Europa's relatively icy leading hemisphere, although this suggestion is based on very limited available laboratory data which demonstrates a decrease in reflectance between 200 and 400 nm for irradiated water ice \citep{sack1991alteration}. For Callisto, it is plausible that such irradiation damaged water ice could contribute a sort of global baseline for the near-UV downturn (below 400 nm), perhaps helping to explain the relatively weak but still prominent near-UV downturn observed in the bright water-ice rich regions. However, the association of the strongest near-UV downturns with dark, water-poor material implies that additional species must be largely responsible for the relative differences in the strength of the near-UV downturn across Callisto's surface.

Radiation processed organics have been invoked to explain the relative strengths of the near-UV downturn in the Saturnian system \citep[e.g.][]{hendrix2018icy}. Organic molecules have long been suggested as a plausible component of the dark material on Callisto \citep[e.g.][]{mccord1997_OrganicsOtherMolecules, mccord1998_NonwatericeConstituentsSurfacea, cartwright2024_RevealingCallistoCarbonrich} and, as previously discussed, various organic species offer a plausible, although unconfirmed, spectral match to several of Callisto's UV and visible wavelength absorption features. It therefore seems likely that organic species contribute to the near-UV downturn on Callisto. Irradiation-produced tholins, as shown in Figure \ref{fig:comp}, provide a close visual match to the broad downturn, especially for wavelengths short-ward of $\sim$400 nm. However, many organic species show deep absorption bands in this wavelength range and multiple species with plausibly overlapping absorptions could also contribute to the observed downturn.

Additional species, such as many silicates and phyllosilicates which have been suggested as plausible explanations for the 450 and 930 nm absorption bands also tend to show strong downturns towards the near-UV (see e.g. serpentinite in Figure \ref{fig:comp}a) and, if present, likely contribute to the observed downturn. Carbonates, which are commonly discussed as a likely component of Callisto's dark material, also display a near-UV downturn and may contribute to the feature on Callisto. Figure \ref{fig:comp} shows the spectrum of shortite, a Na,Ca carbonate (PSF sample CRB126), which matches Callisto's near-UV downturn quite well and is otherwise generally featureless at UV-visible wavelengths. Na- and Ca- bearing carbonates have also been proposed to explain the 4.02 \um band in Callisto's NIR spectrum \citep{johnson2004radiationchapter} and our results, along with the recent analysis of JWST/NIRSpec observations by \citet{cartwright2024_RevealingCallistoCarbonrich}, suggest that Callisto's non-\ch{H2O} spectral properties are dominated by carbon. It therefore seems likely that Callisto's near-UV downturn arises from the combination of multiple species, each with distinct absorptions in the visible and near-UV, perhaps including organics, silicates, and carbonates.

\section{Implications for Callisto's Surface Composition}\label{sec:implications}

The relative contribution of endogenic and exogenic sources to the dark non-icy material which covers much of Callisto's surface is not well constrained. The two most commonly discussed end-member hypotheses are that the dark material is sourced primarily from lag deposits left behind after the ice sublimates out of endogenic ice/rock mixtures that were excavated from Callisto's subsurface by large impactors \citep[e.g.][]{moore1999_MassMovementLandform, moore2004callisto, howard2008_SublimationdrivenErosionCallisto, white2016modeling} or that the dark material is comprised almost entirely of irregular satellite dust that coats the surface in a thick layer and fills in all of the low lying regions \citep[e.g.][]{bottke2013black}. \citet{bottke2013black} argue that the latter is more consistent with observations and suggest that a global layer of irregular satellite dust $\sim$120 - 140 m thick should have built up on Callisto over the last $\sim$4 Gyr. Critically, however, the estimates of \citet{bottke2013black} rely on Callisto's surface age predating irregular satellite capture, as much of the total volume of delivered dust arrives in the first $\sim$40 Myr after the irregular satellites are captured. The slightly younger surface age (3.97-4.15 Ga) for Callisto estimated by \citet{bottke2024bombardment}, combined with the currently favored early instability as part of the Nice model for solar system formation \citep[e.g][]{nesvorny2018evidence, liu2022early} suggests that Callisto's surface may be as much as $\sim$600 Myr younger than the time of irregular satellite capture. Based on the simulations of \citet{bottke2013black}, this could result in an order of magnitude or more decrease in the accumulated height of the irregular satellite dust on Callisto, although it is important to note that uncertainties on both of these timescales are large and the dust delivered before the final global resurfacing event reset Callisto's surface age may still be incorporated into Callisto's subsurface material.

\citet{bottke2013black} argue that the idea of a very thick ($\gtrsim$100m) layer of irregular satellite dust dominating Callisto's dark material is supported by the apparent spectral similarity between the dark material on Ganymede and Callisto and carbonaceous meteorites (CI and CM type) and asteroids (C- and D-type), as noted by multiple authors \citep{denk1998candidate, greeley2000_GalileoViewsGeologya, hibbitts2003impactor}, and based on the expectation that the Jovian irregular satellites may be sourced from the same population as C- and D- type asteroids \citep[e.g.][]{luu1991ccd, bhatt2017composition}. JWST observations have revealed that several of the retrograde irregular satellites do indeed appear to be spectrally consistent with C- and D- type asteroids, although there is an additional unexpected population in the prograde Himalia family that is rich in ammoniated phyllosilicates and likely contributes different material to the dust falling on Callisto \citep{sharkey2025_JWSTRevealsVaried}. However, recent observations with JWST have made it clear that Callisto's dark material is not simply a more icy spectral analog to the irregular satellites, as would be expected for a $\sim$100 m layer of deposited dust that is mixed with relatively pure ice excavated from Callisto's subsurface \citep{cartwright2024_RevealingCallistoCarbonrich, sharkey2025_JWSTRevealsVaried}. Additionally, the albedo of Callisto's darkest material is $\sim$0.2 \citep{moore2004callisto}, while the albedo of the irregular satellite dust is $\sim$0.03-0.06 \citep{grav2015neowise}. It appears unlikely that enough water ice will be excavated and mixed into the dust to increase the albedo from 0.05 to 0.2 if all of the dark material in the 100 m layer is sourced from the irregular satellites, implying that a significant component of (less-dark) endogenic material is likely needed within the lag deposits to explain the observed albedo. The inclusion of significant amounts of dark endogenic material, perhaps rich in silicates or carbon-bearing species, within at least some of Callisto's dark lag deposits, as well as the presence of radiolytically altered species are also required to explain the distinct spatial distributions and regional variations of Callisto's UV and visible spectrum.

Based on our mapped spectral features and the results of other recent observations of Callisto \citep{cartwright2024_RevealingCallistoCarbonrich} and the Jovian irregular satellites \citep{sharkey2025_JWSTRevealsVaried}, we suggest that there are at least three separate compositional regimes within Callisto's dark material which we define as follows:\\
\hspace{10pt} (1) material excavated by, delivered from, or altered by the Valhalla and Asgard impacts which have since been mixed with a relatively small amount of irregular satellite dust and more recently exhumed subsurface material (which itself may have been altered by the giant impact events)\\
\hspace{10pt} (2) material of older surfaces with a relatively larger fraction of irregular satellite dust mixed with endogenic material excavated by large impacts \\
\hspace{10pt} (3) mixtures of endogenic and exogenic material that have been radiolytically altered via magnetospheric bombardment, with the largest affects seen on the trailing hemisphere.

The most prominent UV and visible-wavelength absorption features in our spatially resolved Callisto spectra tend to display unique spatial distributions indicating that different species and material sources are responsible for each of the major absorptions, with the possible exception of the 230 and 450 nm features which may be correlated. \citet{bottke2024bombardment} suggest that impactors only modestly larger than the impactor that formed Valhalla ($\gtrsim$60 km) may have resulted in regional or global erasure events that reset Callisto's surface early in its lifetime. It is therefore plausible that the Asgard and Valhalla impact events may have similarly reset Callisto's surface on a smaller regional scale, erasing previously built up irregular satellite dust and explaining the apparently distinct composition of the regions in and nearby the large impact basins. The terrains within and surrounding Asgard and Valhalla are the only regions with a 320 nm absorption, and there is a clear association between the basins and the 820 and 930 nm bands on Callisto's leading hemisphere. The association of these features with the impact basins implies a possible relationship with subsurface material excavated by the large impactors responsible for the basins, or material that was delivered from or altered by (e.g. via impact shock or melt) these impactors. The unique spatial distribution of the 820 and 930 nm absorptions may be more consistent with creation via impact melt, while the wider reaching 320 nm absorption implies a possible relation to impact ejecta containing material sourced from deep within the subsurface or perhaps from the giant impactors themselves. This explanation may not be consistent with the two patches of material which show relatively strong 820 and 930 nm absorptions on Callisto's trailing hemisphere as there are no known large impact basins nearby. However, these patches are somewhat close to antipodal with Asgard and Valhalla and it is possible that a process similar to antipodal ejecta focusing thought to have occurred for several recent impacts on the moon \citep[e.g.][]{bandfield2017distal, wakita2021impactor}, may have deposited impact ejecta from the Valhalla and Asgard impact events onto Callisto's trailing hemisphere. Recent simulations have shown that antipodal ejecta focusing may be possible on Europa and Ganymede, perhaps even providing enough energy for secondary melt production \citep{gritsevich2023theory}, however, it is unclear whether such models can be applied to Callisto's ancient giant basin-forming impacts, and future modeling work will be required to explore this hypothesis. Regardless, the spatial distributions of the 320, 820 and 930 nm absorption features are inconsistent with the patterns expected for delivery of exogenic material from irregular satellite dust or magnetospheric bombardment, and thus these features are most likely related to material either excavated from Callisto's subsurface or delivered from/altered by the Asgard and Valhalla impact events.

Assuming that the 320 nm absorption band is indeed related to organic species either delivered from, excavated by, or created during the Valhalla and Asgard forming impact events, it may be responsible for the stronger near-UV downturn (redder visible color) generally associated with the impact basins and nearby terrains. Based on the more limited spatial mapping available from JWST, \citet{cartwright2024_RevealingCallistoCarbonrich} find that, like the 320 nm feature, the 4.57 \um band is weaker on the trailing hemisphere. However, unlike the 320 nm feature, the authors find that the 4.57 \um band is strongest at mid-southern latitudes on the anti-Jovian hemisphere (where the 320 nm absorption is relatively weak) and weaker in and around Asgard basin (where the 320 nm feature is stronger). It is clear from their distinctly different spatial distributions that a single species or source of organics cannot be responsible for both the 4.57 \um and 320 nm absorptions. The 4.57 \um band is not seen in the spectra of the irregular satellites \citep{sharkey2025_JWSTRevealsVaried}, and is unlikely to be a simple irradiation product of the dust as it is also not present on Ganymede \citep{bockelee2024composition} which receives the same composition of exogenic dust, just to a lesser degree \citep{bottke2013black}. Taking into account the fact that Callisto maintains a strong $\sim$2-10\% 4.57 \um band depth across its surface, we suggest that the species responsible for this band is most likely formed during irradiation of mixtures of excavated endogenic material and exogenic dust, perhaps favoring a non-organic origin or requiring ammoniated species delivered by the dust as a nitrogen source if the attribution to CN-bearing organics is correct. On the trailing hemisphere, the more intense irradiation environment may result in preferential carbonization of organic species \citep[e.g.][]{hendrix2008_CallistoNewInsights} that produces heavily processed carbon compounds, such as amorphous carbon, with a strong 275 nm absorption edge. This simultaneously explains the presence of the 275 nm absorption edge, absence of the 320 nm band even in the potential antipode deposits, and relative weakness of the 4.57 \um feature on the trailing hemisphere, although some small steady state concentration of the 4.57 \um band material is still produced. This is consistent with the anti-bullseye distribution of the 4.57 \um feature and its apparent anti-correlation with radiolytically produced \ch{CO2}, as noted by \citet{cartwright2024_RevealingCallistoCarbonrich}. On the leading hemisphere, our working model would suggest the regions with the strongest 4.57 \um bands correspond to older terrains which should have the largest fraction of irregular satellite dust mixed with the endogenic material. At Asgard and Valhalla, we suggest that significantly less irregular satellite dust has built up since the formation of the large impact events reset the regional surfaces. While the impacts basins are ancient, perhaps $\sim$2-4 Ga \citep{greeley2000_GalileoViewsGeologya}, the accretion rate of the dust would have been largest early in Callisto's lifetime and most of the dust is expected to have arrived shortly after the capture of the irregular satellites \citep{bottke2013black}. A lower abundance of irregular satellite dust associated with the impact basins may explain the relative weakness of the 4.57 \um absorption on its own, although we cannot rule out the possibility that the altered material present near Asgard and Valhalla may affect the radiolytic chemistry and lead to less efficient formation of the 4.57 \um band.

The 230 nm and 450 nm absorption features may be associated with irradiated \ch{NaCl} on Callisto. If so, then \ch{NaCl} appears to be relatively widespread across the surface. A plausible source of \ch{NaCl} is exogenic delivery associated with the neutral torus from Io. However, we do not see any obvious enhancement in the 450 nm band strength associated with the sub-Jovian quadrant of the leading hemisphere where the neutral torus is expected to deposit the most material \citep{cooper2001energetic}. If the \ch{NaCl} is endogenic, perhaps more in alignment with the generally widespread distribution, it may be exposed from the near sub-surface by impact events and mixed into the ancient regolith, or produced during irradiation of freshly exposed native Na- and Cl- bearing species. It is interesting to note that \ch{NaCl} on Europa is likely to be sputtered into the exosphere over relatively short geologic timescales \citep{johnson2009composition}, requiring a recent source to explain its presence. It is unclear what the lifetime would be on Callisto, and whether the significantly less intense radiation environment alone would allow \ch{NaCl} to persist on its ancient surface. Regardless of its origin, this potential detection of \ch{NaCl} on Callisto along with recent detections on Europa \citep{trumbo2019sodium, trumbo2022new}, a plausible detection of \ch{NaCl} and other hydrated salts in Juno/JIRAM spectra of Ganymede \citep{tosi2024salts}, and the unconfirmed speculation for Na- oxalates (or Na- sulfates) in the JWST Callisto data \citep{cartwright2024_RevealingCallistoCarbonrich}, suggests that Na-bearing salts may be more prevalent on surfaces throughout the Galilean system than previously thought. 

We find little evidence for sulfur-bearing species in Callisto's UV-VIS spectra, perhaps suggesting that sulfur chemistry plays a less significant role in the surface composition than previously suggested \citep[e.g.][]{noll1997_DetectionSO2Callisto, mccord1997_OrganicsOtherMolecules, hibbitts2000_DistributionsCO2SO2, cartwright2020_EvidenceSulfurbearingSpecies}. The 280 nm ratio band does not require an additional absorber, like \ch{SO2}, to explain its presence or spatial distribution and is better explained by the presence of the 275 nm absorption edge on the trailing hemisphere and the 320 nm band on the leading hemisphere. \citet{cartwright2020_EvidenceSulfurbearingSpecies} suggest that thermally altered sulfur allotropes or \ch{HS2} may explain the 4.02 \um absorption which is stronger on the leading hemisphere and discuss various plausible sources of sulfur at Callisto \citep[see][and references therein]{cartwright2020_EvidenceSulfurbearingSpecies}. However, these species are expected to show strong UV absorptions \citep{gosavi1973uv, gradie1980surface, moses1991phase}, which we do not see in our HST observations, likely ruling out a significant contribution from sulfur allotropes or \ch{HS2} to Callisto's NIR spectrum. The irregular satellites, whose dust has been suggested as a source of sulfur to Callisto's leading hemisphere, show no evidence of sulfur-bearing species in recent JWST observations \citep{sharkey2025_JWSTRevealsVaried}. If the 820 and 930 nm bands are associated with sulfate species, a hypothesis we generally disfavor, this could be consistent with an endogenic source of sulfur. However, these absorptions are likely associated with impact altered material, some of which may have been delivered directly from the giant impactors themselves, and are not widespread across Callisto's surface. These bands are therefore unlikely to be related to the strong NIR features, which may or may not be associated with sulfur-bearing species.

While we do not see clear spectroscopic evidence of sulfur-bearing species on Callisto's surface, some amount of sulfur is expected to be delivered by the neutral torus and magnetospheric ion bombardment. Existing simulations and Galileo fly-by measurements suggest that Iogenic \ch{S^n} ions should predominantly impact the trailing hemispheres of the icy Galilean satellites \citep{cooper2001energetic}. Suggestions in the literature that this pattern may be reversed at Callisto \citep[e.g.][]{cartwright2020_EvidenceSulfurbearingSpecies} are largely based on the presumed presence of \ch{SO2} and other sulfur bearing species on the leading hemisphere, although recent simulations have shown that under certain conditions the impact pattern of high energy sulfur ions may shift away from the trailing hemisphere and, in some extreme cases, preferentially bombard the leading hemisphere \citep{liuzzo2019energetic}. We find little observational motivation to support (or refute) any of these proposed sulfur bombardment patterns. The sulfur bombardment rate at Callisto, which is 2-3 orders of magnitude lower than at Europa \citep{cooper2001energetic}, may be too low to drive significant radiolytic sulfur chemistry, although the delivered sulfur must be incorporated somehow as we do not find any evidence of sulfur allotropes which typically have strong UV absorptions. Perhaps the radiolytically produced sulfur species do not strongly affect Callisto's UV-VIS spectrum or the sulfur is incorporated into the so-called carbon cycle and contribute to the UV spectral features which we suspect may be related to organic species. Alternatively, if most of the sulfur is converted to hydrated sulfuric acid (\ch{H2SO4}), as on Europa's trailing hemisphere \citep{carlson1999sulfuric, carlson2002sulfuric}, and which has recently been suggested to explain a prominent 5.9 \um band on Ganymede \citep{bockelee2024composition}, then it may not significantly affect our observed HST spectrum as the $\sim$190 nm electronic transition band of sulfuric acid \citep{zhou2022so3} is outside of our observed spectral range. Future observations, especially at far-UV and mid-IR wavelengths, will be necessary to constrain the ultimate fate of magnetospheric sulfur at Callisto.

\section{Comparison Between Galilean Satellites}\label{sec:comparison}

\begin{figure*}[ht!]
\includegraphics[width=\textwidth]{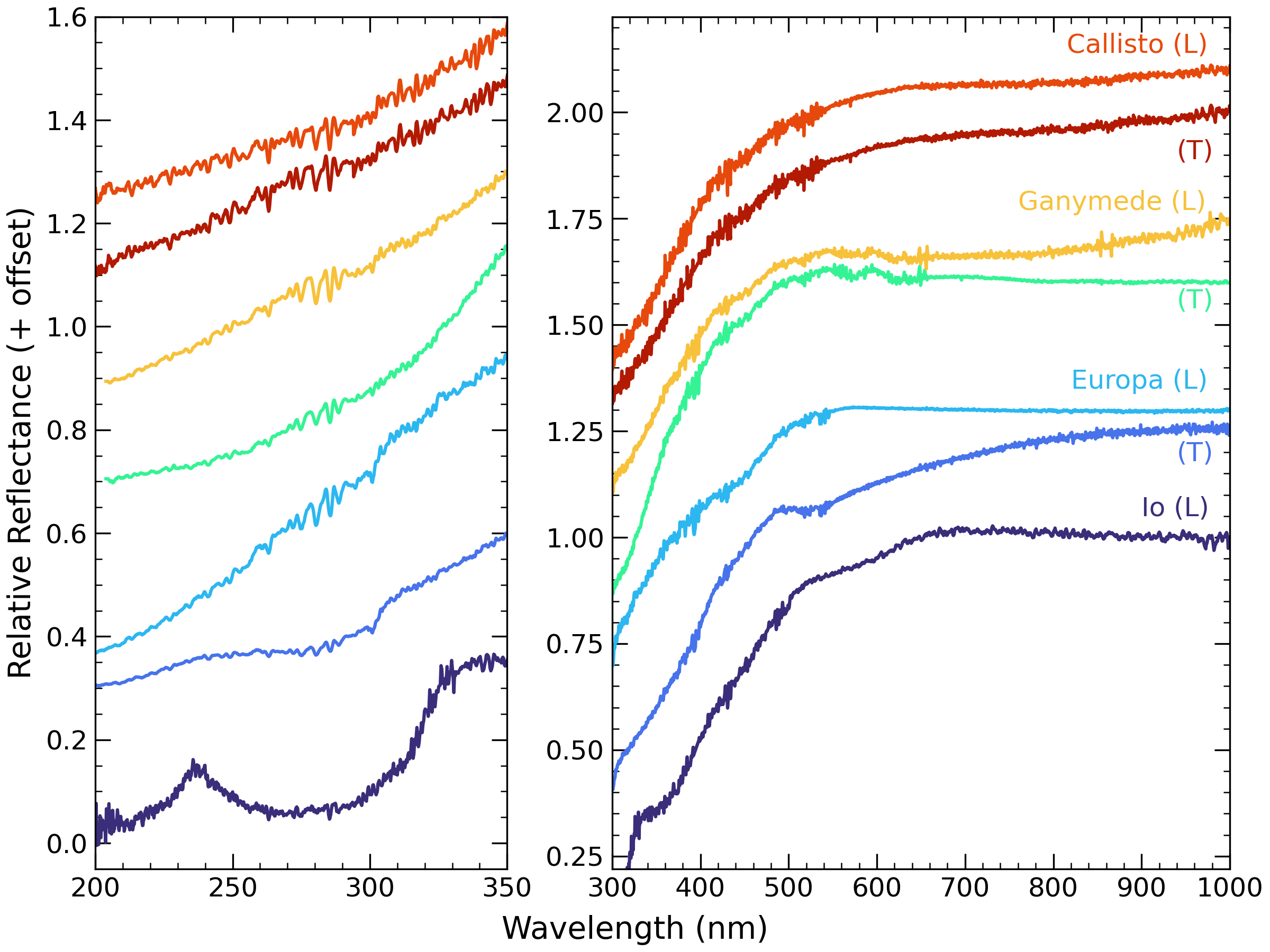}
\caption{Comparison of HST/STIS UV (left) and visible (right) spectra of the leading and trailing hemispheres of Callisto, Ganymede, and Europa, and a patch of bright white (sulfur rich) material on Io's leading hemisphere. The leading and trailing averages for Europa, Ganymede, and Callisto include all UV and visible pixels within $\sim$53 degrees of the leading and trailing points, respectively, combined to create a full 200 - 1000 nm spectrum. All spectra are normalized to 1000 nm, and offset for clarity (+0, +0.25, +0.3, +0.6, +0.75, +1, and +1.1).}
\label{fig:comparison}
\end{figure*}

A comparison between the UV and visible spectra of the four Galilean satellites from HST/STIS is shown in Figure \ref{fig:comparison}. The Io spectrum is from a patch of bright white sulfur-rich material on the leading hemisphere \citep{trumbo2022_SpectroscopicMappingIo}. Average spectra of the leading and trailing hemispheres of the three icy satellites - Europa, Ganymede, and Callisto are also shown \citep{trumbo2019sodium, trumbo2020endogenic, becker2022_MidultravioletHubbleObservations, trumbo2021_GeographicDistributionDensephase}. All of the hemispherically averaged satellite spectra have red slopes in the near-UV (200 - 350 nm), a steep red slope from $\sim$400 - 500 nm, and a relatively flat or slightly red slope from $\sim$700 - 1000 nm. All of the satellite spectra also show a strong downturn toward the near-UV, with a band edge near 500 nm, consistent with previous ground-based observations \citep[e.g.][]{mcfadden1980_VisibleSpectralReflectance, calvin1995_SpectraIcyGalileanc}. This near-UV downturn appears to be somewhat more pronounced in the spectrum of Io and on Europa's trailing hemisphere than on Europa's leading hemisphere or on Ganymede and Callisto, likely due to an additional contribution from sulfur allotropes at Io and on Europa's trailing side \citep[e.g.][]{spencer1995charge,trumbo2022_SpectroscopicMappingIo, trumbo2020endogenic}. However, this near-UV downturn is ubiquitous across all of the Galilean satellites, as well as the icy Saturnian satellites \citep{hendrix2018icy}, so sulfur cannot be the sole cause for all of these bodies. \citet{hendrix2018icy} suggest radiation processed organics may explain the relative strengths of the near-UV downturn in the Saturnian system. As discussed in Section \ref{sec:labcomp_nearUV}, it is possible that irradiation processed organics could also contribute to the near-UV downturn in the Galilean system, especially on Ganymede and Callisto. The presence of hydrated silicate minerals on Ganymede and Callisto could also contribute to the near-UV downturn, as has been proposed to explain the similar near-UV downturn short-wards of $\sim$400 nm on hydrous meteorites and primitive asteroids \citep[e.g.][]{gaffey1978asteroid, cloutis2011spectralCI, cloutis2011spectralCM}. Indeed, both hemisphere spectra of Ganymede and Callisto have an apparent slope break near 400 nm, which is consistent with a contribution from multiple species to their near-UV downturn. Perhaps the presence of hydrated iron-bearing silicates within the dark material of both moons contributes to a deepening of the near-UV downturn short-wards of $\sim$400 nm. \citet{trumbo2020endogenic} has also suggested irradiation-damaged water as a potential explanation for the near-UV downturn on Europa's leading hemisphere, which is weaker than the trailing hemisphere where sulfur is thought to contribute. Irradiation damaged water ice could contribute to the UV absorption for all of the icy Galilean satellites, although it is likely to have the strongest contribution for Europa's leading hemisphere, where comparatively pure water ice is most abundant.

As noted in \citet{trumbo2022_SpectroscopicMappingIo}, the Io spectrum in Figure \ref{fig:comparison} has a strong UV \ch{SO2} signature, with reflectance minima near 220 and 280 nm and a local maximum near 240 nm \citep{nash1980so2}. This \ch{SO2} signature is also clearly seen in the trailing hemisphere spectrum of Europa, albeit more muted, and is attributed to irradiation-produced \ch{SO2} from Iogenic sulfur ions deposited onto Europa's trailing hemisphere by Jupiter's magnetosphere \citep[e.g.][]{lane1981evidence, hendrix2011europa, becker2022_MidultravioletHubbleObservations}.  Despite past claims of a similar \ch{SO2} band in Callisto's leading/trailing spectral ratio \citep{noll1997_DetectionSO2Callisto}, our analysis suggests that this feature and identification are spurious and, in reality, the consequence of dividing the unrelated 320 nm leading hemisphere band by the trailing-hemisphere 275 nm absorption edge. There is no direct evidence of a UV \ch{SO2} signature on either hemisphere of Ganymede or Callisto.

There is also a broad absorption from $\sim$200 - 360 nm that can be seen in Ganymede's trailing hemisphere spectrum in Figure \ref{fig:comparison}. \citet{noll1996detection} took the ratio of Ganymede's disk integrated trailing hemisphere to the leading, and attributed the observed $\sim$260 nm feature to the \ch{O3} Hartley band. Future studies noted that Ganymede's absorption is broadened relative to that of \ch{O3}, and an additional component attributed to \ch{SO3-} and \ch{HSO3-} radicals produced during sulfur implantation into oxygen rich water ice may further improve the fit to Ganymede's 260 nm absorption on the short-wavelength side \citep{boduch2016ion}. An additional absorption near $\sim$300 nm from an unidentified UV absorber may also be required to fit the long-wavelength side of Ganymede's 260 nm feature \citep{teolis2006ozone, jones2014uv}. We note, however, that these studies relied on data which only extended to $\sim$330 nm and our HST spectra show that the broadening of Ganymede's 260 nm absorption on the long-wavelength side actually extends out to $\sim$360 nm (see Figure \ref{fig:comp_280nm} for a ratio of Ganymede's trailing to leading hemisphere), which places the band center of this additional UV absorber closer to 320 nm. This absorption is apparent in the average trailing hemisphere spectrum of Ganymede and appears to be a close spectroscopic match to the 320 nm feature seen primarily on Callisto's leading hemisphere which we tentatively attribute to the presence of organics. We note, however, that it is difficult to determine where the short end of Ganymede's $\sim$320 nm absorption is because it blends into the strong \ch{O3} band, and confirmation that the feature seen on Ganymede matches Callisto's 320 nm band will require careful analysis of Ganymede's disk-resolved UV-visible spectra, which is beyond the scope of this paper. Nonetheless, it is worth considering whether the species responsible for the 320 nm absorption may be present on both Ganymede and Callisto, and that this could be the unidentified UV absorber required to fit Ganymede's broadened 260 nm feature. 

Europa and Ganymede's leading hemisphere and Callisto's trailing hemisphere spectra show a drop in reflectance from $\sim$275 nm to 200 nm which has previously been attributed to \ch{H2O2} \citep{hendrix1999_HYDROGENPEROXIDEICY}. However, the 3.5 \um \ infrared peroxide band has only been identified on Europa's leading hemisphere and at high latitudes on Ganymede \citep{trumbo2019_H2O2ChaosTerrain, trumbo2023_HydrogenPeroxideGanymede, wu2024europa}. As seen in Figure \ref{fig:comparison}, Ganymede and Callisto have very similar UV spectral slopes, and the generally equatorial distribution of the UV red material on Ganymede is inconsistent with an origin of \ch{H2O2}. Instead, organics may dominate the reddening of the UV spectral slope at Ganymede and Callisto, as suggested by \citet{hendrix2008_CallistoNewInsights}. For Europa, \ch{H2O2} likely contributes to the reddening of the UV slope on the leading hemisphere, although an additional contribution from organics or other species cannot be ruled out.

There is a clear 577 nm absorption due to condensed \ch{O2} in the trailing hemisphere spectrum of Ganymede, which is expected as Ganymede's \ch{O2} is concentrated at the low- to mid-latitudes of the trailing hemisphere \citep{trumbo2021_GeographicDistributionDensephase} and the disk-averaged trailing hemisphere absorption is $\sim$2\%. A weaker absorption is also seen in the Ganymede leading hemisphere spectrum. The 577 nm \ch{O2} absorption is not seen in the Europa or Callisto spectrum in Figure \ref{fig:comp}, however this is not surprising as ground-based detection suggests the 577 nm absorption is only $\sim$0.3\% deep on Europa and Callisto \citep{spencer2002_CondensedO2Europa}---at the level of the noise for these HST/STIS observations.

The Europa leading hemisphere spectrum has a broad absorption centered at 450 nm (350 to 530 nm), and a weaker UV absorption at 230 nm (200 to 260 nm) which has been attributed to irradiated \ch{NaCl} in the leading hemisphere chaos terrains \citep{trumbo2019sodium, trumbo2022new}. Both the leading and trailing hemisphere spectra of Ganymede and Callisto show a weak absorption feature centered near 450 nm, which appear to be a match to the band center of Europa's 450 nm absorption, although it is difficult to determine the width and shape of the feature on either satellite due to the quickly changing continuum at the surrounding wavelengths due, at least in part, to the near-UV downturn. Additionally, a broad but weak absorption centered near 230 nm can be seen in the trailing hemisphere spectra of both Ganymede and Callisto, and plausibly, but less obviously, in the leading hemisphere spectra as well. We find that Callisto's 230 nm and 450 nm absorptions may be correlated with one another, although difficulties with mapping these features make confirmations of this impossible. Nonetheless, Callisto's absorptions are plausibly consistent with the laboratory spectra of irradiated \ch{NaCl} and features detected on Europa, providing the first reported spectral evidence for the presence of \ch{NaCl} on Callisto. Based on these hemispherically averaged spectra, \ch{NaCl} is also a likely spectroscopic match to Ganymede's 230 nm and 450 nm features, although confirmation of this will require careful analysis of the full Ganymede data set and verification that the absorptions are spatially correlated across the surface of Ganymede as well, which is beyond the scope of this paper. However, the presence of \ch{NaCl} on Ganymede is consistent with the recent plausible detection of \ch{NaCl} and other hydrated salts in Juno/JIRAM spectra of Ganymede \citep{tosi2024salts}.

There is a 360 nm absorption seen in the trailing hemisphere Europa spectrum which is interpreted to result from irradiation induced alteration of an unidentified endogenous species \citep{trumbo2020endogenic}. This feature does not appear in any of the other satellite spectra, or in Europa's leading hemisphere spectrum. Additionally, a 530 nm absorption in Europa's trailing hemisphere spectrum and 560 nm absorption in the Io spectrum, commonly attributed to the presence of sulfur allotropes \citep[.g.][]{spencer1997volcanic,  carlson2009_EuropaSurfaceComp, trumbo2020endogenic, trumbo2022_SpectroscopicMappingIo}, are not present in any of the Ganymede or Callisto spectra or on Europa's leading hemisphere.

\section{Conclusion}

We present global, spatially resolved 200 - 1000 nm spectra of Callisto obtained with HST/STIS. We use this dataset to map observed spectral features and explore possible compositions of Callisto’s surface material through comparison with laboratory samples and in the context of alteration processes known to affect Callisto's surface chemistry. We map the strength of a widespread downturn toward the near-UV and the short-wavelength NIR spectral slope from 700 to 1000 nm, which varies from slightly blue to red across Callisto’s surface. At a global scale, bright water-ice-rich regions tend to have neutral or blue NIR slopes and a shallower near-UV downturn, while darker material is generally associated with red NIR slopes and stronger near-UV absorption. The strength of the 700 - 1000 nm slope does not correlate strongly with albedo differences in the dark material, but rather varies regionally with the NIR reddest slopes associated with dark material east of the Valhalla impact basin. Consistent with the hemispherical and regional color variations of dark material noted in SSI imaging, we find that global correlations between absorption features and albedo are generally weak on Callisto, but become stronger when examined regionally, suggesting that the composition of Callisto’s non-ice material varies on regional scales. We find that there are at least three distinct compositional regimes for Callisto's dark material, including (1) material excavated by, delivered from, or altered by the Valhalla and Asgard impacts which have since been mixed with a relatively small amount of irregular satellite dust and exhumed subsurface material, (2) older surfaces with a relatively larger fraction of irregular satellite dust mixed with endogenic material excavated by impacts, and (3) mixtures of endogenic and exogenic material that have been radiolytically altered via magnetospheric bombardment, with the largest effects generally seen on the trailing hemisphere.

Maps of the overlapping absorption bands at 820 nm and 930 nm, which are generally found together in varying ratios, suggest a strong correlation with the Asgard and Valhalla impact basins and may be consistent with an origin from impact melt. The 930 nm absorption band may be related to an Fe-bearing silicate, most plausibly \ch{Fe^{2+}}-rich phyllosilicate or a mixture of Ca-poor orthopyroxene and CM2-like phyllosilicates, or Na-bearing sulfate. The 820 nm band can possibly be explained by either a currently unidentified silicate or ferrous ammonium sulfate. Two large regions of dark material on the trailing hemisphere, not associated with any known giant impact basin, show a strong 930 nm absorption with a relatively weak 820 nm feature. We suggest that these patches may be associated with near-antipodal deposits of material from the Asgard and Valhalla impact events, although significant modeling work and a more complete understanding of the extent and composition of these patches will be necessary to confirm this hypothesis. We also map a 320 nm absorption feature which is a plausible spectral match to various unsaturated organic species and is only present in regions within and surrounding the Asgard and Valhalla basins, perhaps hinting at an association with impact ejecta containing material from deep within the subsurface or perhaps from the giant impactors themselves. We suspect that the impact events which formed these basins may have regionally reset Callisto's surface composition, burying any irregular satellite dust which had previously been deposited in this region and resulting in comparatively less dust having accumulated near the giant basins than elsewhere on Callisto's surface. We also find that irradiation processed and carbonized organics provide a more consistent explanation for the strong 275 nm absorption edge seen on Callisto's trailing hemisphere, which had previously been attributed to \ch{H2O2}. 

We report two new absorption features near 450 and 230 nm, which may be spatially correlated, and if so are plausibly attributed to irradiated \ch{NaCl}. The apparently widespread distribution of these absorptions is generally inconsistent with the delivery of Iogenic \ch{NaCl} from the neutral torus, which should preferentially bombard the sub-Jovian quadrant of the leading hemisphere \citep{cooper2001energetic}. It is possible that Iogenic \ch{NaCl} may be sputter re-deposited across Callisto's surface or that the \ch{NaCl} is sourced from endogenic material excavated by impacts. Future work exploring the spatial distribution and confirming a correlation between the 230 nm and 450 nm absorption bands is necessary to confirm the presence of \ch{NaCl} on Callisto. Other possible attributions include an Fe-bearing phyllosilicate like serpentinite for the 450 nm band, and various organics for either band.

We find that a 280 nm band seen only in spectra ratioed against the trailing hemisphere and previously attributed to \ch{SO2} \citep{noll1997_DetectionSO2Callisto} is better explained by the presence of the 275 nm absorption edge on the trailing hemisphere and the 320 nm band which is constrained to the leading hemisphere. The presence of an additional absorber with a 280 nm band center on the leading hemisphere, like \ch{SO2}, is not required to explain this ratio band or its spatial distribution. Generally, we do not find evidence of sulfur-bearing species at UV-VIS wavelengths and suggest that the main source of sulfur at Callisto may be magnetospheric \ch{S^n} ions. The sulfur bombardment rate at Callisto, which is 2-3 orders of magnitude lower than at Europa \citep{cooper2001energetic}, may be too low to drive significant radiolytic sulfur chemistry, or perhaps the sulfur is incorporated into organic molecules. Alternatively, existing sulfur species may not strongly affect the UV-VIS spectrum above 200 nm as might be expected for hydrated sulfuric acid.

\newpage
\section*{acknowledgments}
This work is based on observations made with the NASA/ESA Hubble Space Telescope, obtained at the Space Telescope Science Institute, which is operated by the Association of Universities for Research in Astronomy, Inc., under NASA contract NAS526555. These observations are associated with program \#15925. The specific observations analyzed in this paper can be accessed via \dataset[doi: 10.17909/q7h0-3j07]{https://doi.org/10.17909/q7h0-3j07}. Support for program \#15925 was provided by NASA through a grant from the Space Telescope Science Institute, which is operated by the Association of Universities for Research in Astronomy, Inc., under NASA contract NAS526555. The authors thank
the anonymous referees for insightful comments and suggestions that led to an improved manuscript. MRD thanks Dr. Bethany Ehlmann, Dr. William Denmann, Dr. Bill Bottke, Abigail Keebler, Maria Camarca, Swaroop Chandra, Merritt McDowell, and Ashma Pandya for helpful discussions.

\vspace{5mm}
\facilities{HST(STIS)}

\bibliography{bibliography}{}
\bibliographystyle{aasjournal}

\end{document}